\documentclass[pra,aps,longbibliography,twocolumn,superscriptaddress,floatfix]{revtex4-1}
\usepackage{amsmath,amssymb,graphicx,units}
\usepackage[usenames,dvipsnames]{color}
\usepackage{verbatim}
\usepackage{lipsum}
\usepackage{setspace}
\usepackage{float}
\usepackage{dcolumn}
\usepackage{booktabs}
\usepackage{bm}
\usepackage{multirow}
\usepackage[bookmarks=false,linkcolor=blue,urlcolor=blue,colorlinks,citecolor=blue]{hyperref}

\usepackage{mathrsfs}
\usepackage[T1]{fontenc}

\begin{document}

\title{Entanglement growth and information capacity in a quasiperiodic system with a single-particle mobility edge}

\author{Yuqi Qing}
\affiliation{Institute of Physics, Chinese Academy of Sciences, Beijing 100190, China}
\affiliation{Tsung-Dao Lee Institute, Shanghai Jiao Tong University, Shanghai 201210, China}

\author{Yu-Qin Chen}
\email{yqchen@gscaep.ac.cn}
\affiliation{Graduate School of China Academy of Engineering Physics, Beijing 100193, China}

\author{Shi-Xin Zhang}
\email{shixinzhang@iphy.ac.cn}
\affiliation{Institute of Physics, Chinese Academy of Sciences, Beijing 100190, China}

\date{\today}

\begin{abstract}
We investigate the quantum dynamics of a one-dimensional quasiperiodic system featuring a single-particle mobility edge (SPME), described by the generalized Aubry-Andr\'e model. This model offers a unique platform to study the consequences of coexisting localized and extended eigenstates, which contrasts sharply with the abrupt localization transition in the standard Aubry-André model. We analyze the system's response to a quantum quench through two complementary probes, entanglement entropy (EE) and subsystem information capacity (SIC). We find that the SPME induces a smooth crossover in all dynamical signatures. The EE saturation value exhibits a persistent volume-law scaling in the mobility-edge phase, with an entropy that continuously decreases as the number of available extended states decreases. Complementing this, the SIC profile interpolates between the linear ramp characteristic of extended systems and the information trapping behavior of localized ones, directly visualizing the mixed nature of the underlying spectrum. Our results establish unambiguous dynamical fingerprints of a mobility edge, providing a crucial non-interacting benchmark for understanding information and entanglement dynamics in more complex systems with mixed phases.
\end{abstract}

\maketitle


\textit{Introduction.---}
Understanding thermalization and its breakdown in isolated quantum systems is a fundamental problem in quantum many-body physics~\cite{dalessio2016quantum, deutsch2018eigenstate}. While generic interacting systems typically thermalize following the eigenstate thermalization hypothesis~\cite{deutsch1991quantum, srednicki1994chaos, rigol2008thermalization}, this behavior can be subverted by strong quenched disorder. For example, quenched disorder can cause Anderson localization in non-interacting systems~\cite{anderson1958absence} and many-body localization (MBL) can emerge when interaction is turned on, leading to a nonergodic phase where memory of the initial state persists~\cite{fleishman1980interactions, basko2006metal, nandkishore2015manybody, abanin2019colloquium}.

Besides random disorder, quasiperiodic potentials~\cite{thouless1983quantization, han1994critical, iyer2013manybody, schreiber2015observation, lee2017manybody, khemani2017two, chandran2017localization, setiawan2017transport, zhang2018universal, zhang2019strong, wang2020onedimensional, zhai2020manybody} offer an alternative path to localization. The standard Aubry-Andr\'e (AA) model, for instance, shows a sharp transition where all single-particle states become localized at the critical potential strength~\cite{aubry1980analyticity, harper1955single}. A more intriguing scenario is presented by the generalized Aubry-Andr\'e (GAA) model~\cite{biddle2010predicted, ganeshan2015nearest}, which holds a more complex spectral structure with a single-particle mobility edge (SPME)—an energy threshold separating coexisting localized and extended eigenstates. The recent experimental realization of such models in ultracold atoms~\cite{luschen2018singleparticle, kohlert2019observation} and photonic lattices~\cite{verbin2015topological} have provided excellent platforms for theoretical study. Understanding the precise dynamical signatures in the non-interacting setting with SPME is a critical first step, providing an essential baseline for tackling the more complex and debated questions of thermalization in interacting systems that may host a mobility edge~\cite{li2015manybody, modak2015manybody, li2016quantum, deng2017manybody, modak2018criterion, huang2023incommensurate}.

Concepts from quantum information theory offer powerful tools for probing quantum many-body systems~\cite{amico2008entanglement, eisert2010colloquium, abanin2019colloquium}. Among these, entanglement entropy (EE) is a primary probe of correlations and quantum phase structures. The scaling of EE with subsystem size clearly separates different quantum phases, for example, ground states of gapped systems follow an area law. In terms of time-evolved states, Anderson localization systems follow an area law, while thermalizing systems follow a volume law~\cite{bauer2013area, kaufman2016quantum}. Beyond its late-time saturation value, the initial growth of EE after a quantum quench serves as a powerful dynamical probe. Linear growth is a hallmark of chaotic thermalizing systems, logarithmic growth is characteristic of MBL, and a near-complete lack of growth signifies Anderson localization~\cite{bardarson2012unbounded, nanduri2014entanglement, lewis-swan2019dynamics}. Although previous theoretical works have focused on static spectral properties or eigenstate of GAA model \cite{li2016quantum}, it remains an open question how these spectral features translate into observable non-equilibrium dynamics starting from a product state. This raises a key question for the GAA model regarding how does entanglement evolve in a system defined by the coexistence of localized and extended modes? Does it obey a volume law due to the extended states, an area law governed by the localized ones, or does a new, intermediate behavior emerge?

While EE quantifies the overall magnitude of entanglement, it does not fully resolve how quantum information propagates and where it is stored. Other tools reveal different aspects of information dynamics. For example, mutual information tracks shared correlations~\cite{wolf2008area}, and its extensions like tripartite mutual information can detect delocalized entanglement and information scrambling~\cite{hosur2016chaos}. Moreover, out-of-time-ordered correlators are frequently used to diagnose the spreading of local information into non-local correlations—a signature of quantum chaos~\cite{maldacena2016bound, fan2017outoftimeorder, nahum2018operator}. However, to obtain a more direct, spatially resolved picture of how information propagates and is retained in space-time, we employ the subsystem information capacity (SIC)~\cite{chen2025subsystem}, a recently introduced probe of information propagation and retention. This quantity investigates the information flow by viewing a subsystem's evolution as an effective quantum channel whose capacity is shaped by the system's overall dynamics~\cite{holevo2012quantum}. Defined as the mutual information \(I(A{:}R)\) between an output subsystem \(A\) and a reference system \(R\) initially entangled with an input region \(E\), SIC measures how much of the initial quantum information can be recovered from \(A\) after some time. This measure is closely related to the quantum coherent information of the effective channel \(\mathcal{E}_{E \to A}\)~\cite{schumacher1996quantum, chen2025subsystem}, reflecting the system's ability to preserve quantum information and acting as a measure of the channel's single-shot quantum capacity~\cite{lloyd1997capacity, devetak2005distillation}. Importantly, SIC shows distinct spatial profiles in different dynamical phases—from information trapping behavior in localized systems to linear growth in extended ones~\cite{chen2025subsystem}. This makes SIC an ideal diagnostic for the mixed dynamics expected in the GAA model, allowing us to ask whether the contributions of localized and extended modes can be spatially separated.

In this work, we employ two complementary quantum information probes—entanglement entropy and subsystem information capacity—to conduct a comprehensive analysis of the quench dynamics in the non-interacting GAA model. By contrasting its behavior with the abrupt localization transition in the standard AA model, we establish clear and unambiguous dynamical fingerprints of an SPME. We find that the SPME induces a crossover in both dynamical signatures, replacing the sharp transition. Specifically, the EE saturation value exhibits a persistent volume-law scaling, yet with a continuously tunable entropy controlled by the number of available extended states. Complementing this, the SIC profile directly visualizes the mixed spectral nature, interpolating between the linear ramp characteristic of extended systems and the information trapping of localized ones. Our results provide a definitive non-interacting benchmark crucial for understanding information and entanglement dynamics in interacting systems with mixed phases.


\textit{Model and Methods.---}
We study the dynamics of noninteracting spinless fermions on a one-dimensional lattice with open boundary conditions, described by the GAA Hamiltonian
\begin{equation}
H = -t \sum_{i=1}^{L-1} (c_i^\dagger c_{i+1} + \text{h.c.}) + \sum_{i=1}^L \mu_i\, c_i^\dagger c_i, \label{eq:hamiltonian}
\end{equation}
where \(c_i^\dagger\) (\(c_i\)) creates (annihilates) a fermion at site \(i\), \(L\) is the system size, and \(t\) is the nearest-neighbor hopping, set to unity (\(t=1\)). The on-site quasiperiodic potential is
\begin{equation}
\mu_i = 2\lambda \frac{\cos(2\pi b i + \phi)}{1 - a \cos(2\pi b i + \phi)}, \label{eq:potential}
\end{equation}
with \(\lambda\) controlling the potential strength, \(b = (\sqrt{5} - 1)/2\) an irrational number, \(\phi=0\) a global phase, and \(a\) a parameter that deforms the standard AA model (recovered at \(a = 0\)). All calculations are performed at the half-filling sector using the quantum software {\sf TensorCircuit-NG}~\cite{*[{ }] [{. \url{https://github.com/tensorcircuit/tensorcircuit-ng}.}] Zhang2023tensorcircuit}.

\begin{figure}
    \centering
    \includegraphics[width=0.9\linewidth]{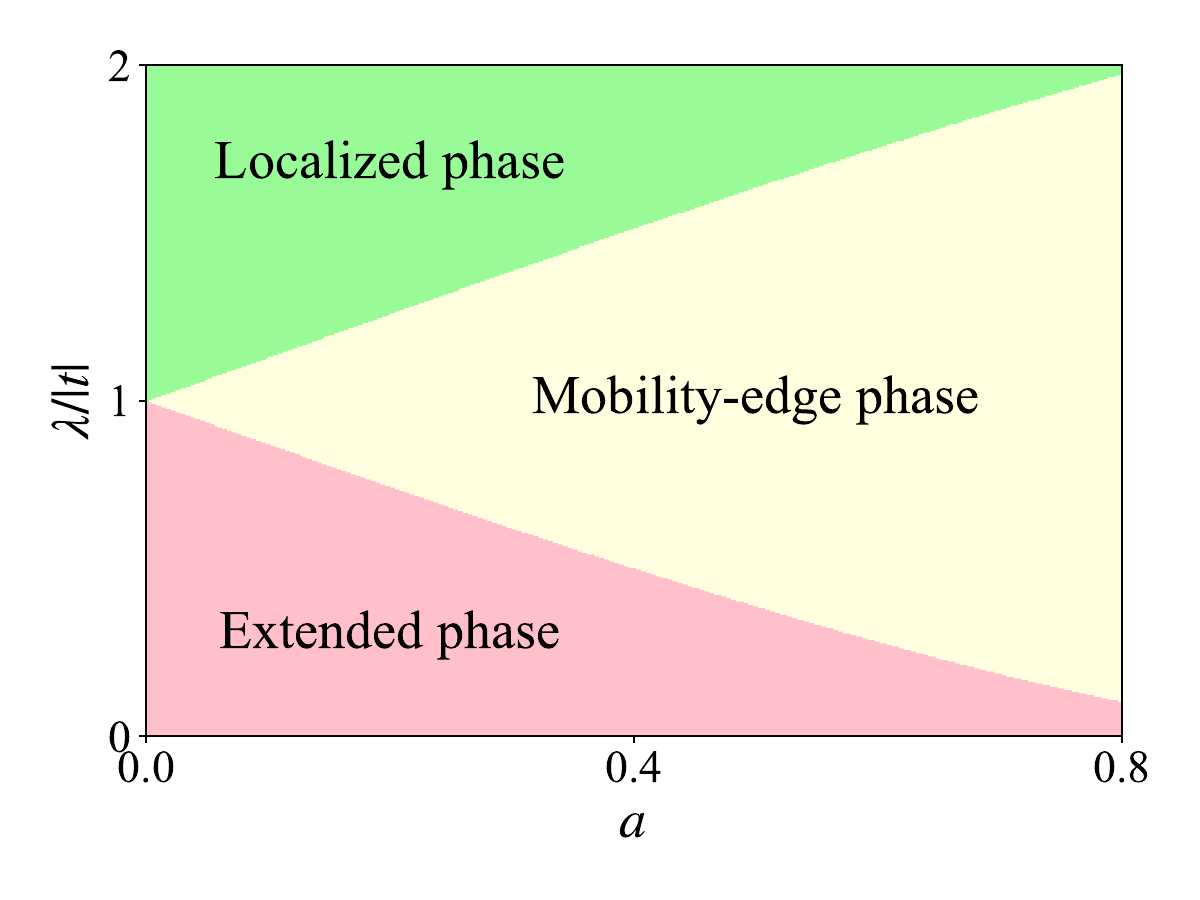}
    \caption{Phase diagram of the GAA model, featuring three distinct dynamical regimes based on the potential strength \(\lambda/|t|\) and deformation \(a\). The model has completely extended (pink) and localized (green) phases, separated by an intermediate phase (yellow) where extended and localized states coexist due to the SPME.}
    \label{fig1}
\end{figure}

The standard AA model (\(a=0\)) features a sharp localization transition at \(\lambda = t\), where all single-particle eigenstates abruptly transition from being extended to localized~\cite{aubry1980analyticity}. For \(a \neq 0\), the GAA model hosts a single-particle mobility edge at energy \(E_c\), given by~\cite{ganeshan2015nearest}
\begin{equation}
    a E_c = 2\, \mathrm{sgn}(\lambda)(|t|-|\lambda|). \label{eq:Ec}
\end{equation}
This critical energy leads to the phase diagram in Fig.~\ref{fig1}, showing a completely extended, a completely localized, and an intermediate phase with a mobility edge where extended and localized states coexist.

\begin{figure}
    \centering
    \includegraphics[width=1\linewidth]{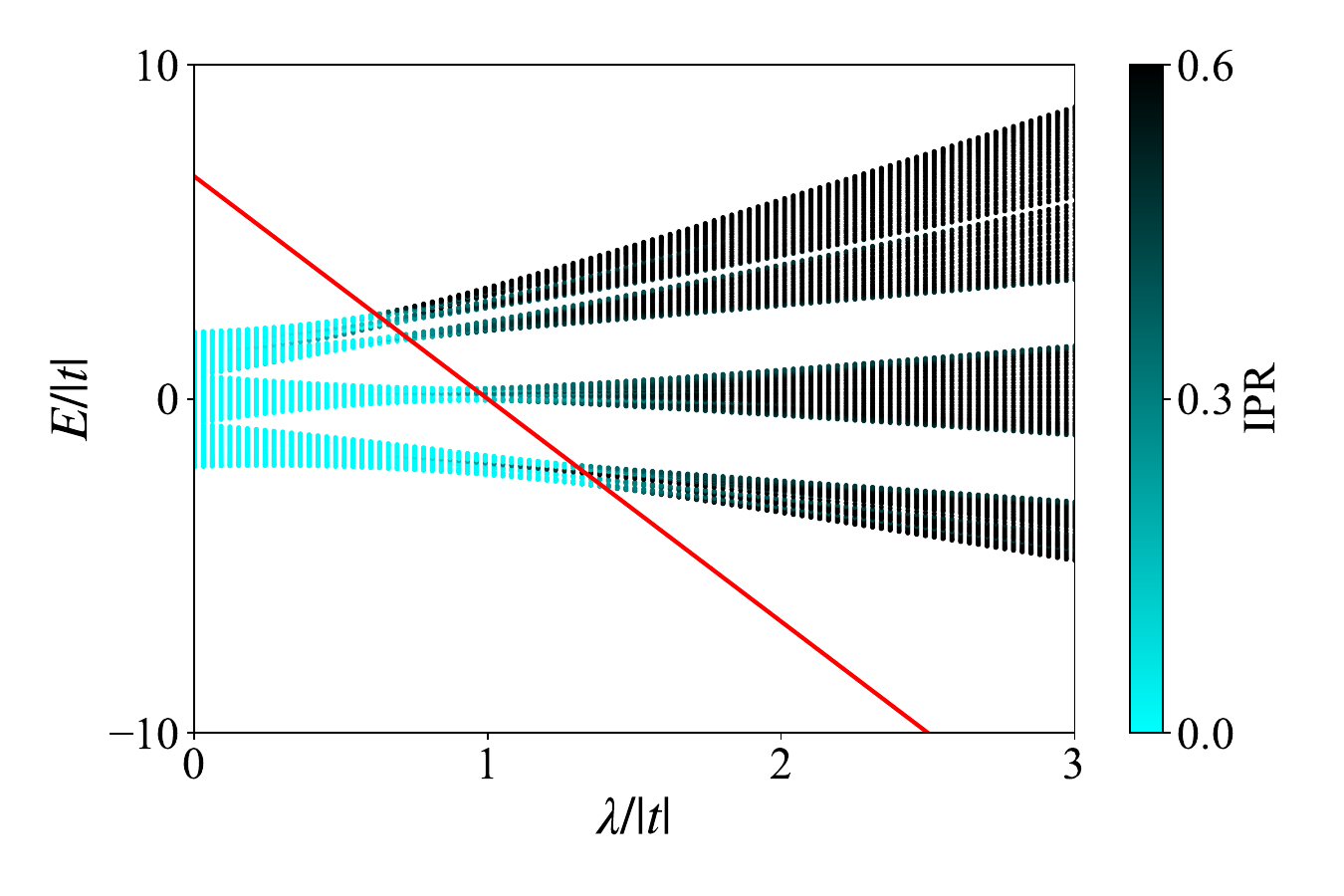}
    \caption{Energy spectrum of the GAA model for \(a=0.3\) and \(L=200\), with eigenstates colored by their IPR. The mobility edge \(E_c\) (red lines), from Eq.~(\ref{eq:Ec}), clearly separates extended states (low IPR, bright cyan) from localized states (high IPR, dark teal/black). This energy-dependent separation is a key feature of the SPME phase.}
    \label{fig2}
\end{figure}

The key feature of the GAA model for \(a \neq 0\) is the coexistence of localized and extended states. The effect of the SPME is visualized by the inverse participation ratio (IPR) of the eigenstates,
\begin{equation}
    \mathrm{IPR}_n = \frac{\sum_{i=1}^{L} |\psi_{n,i}|^4}{(\sum_{i=1}^{L} |\psi_{n,i}|^2)^2}, \label{eq:IPR}
\end{equation}
where \(\psi_{n}\) is the \(n\)-th eigenstate of the system. As shown in Fig.~\ref{fig2} for \(a=0.3\), the eigenstates are clearly separated into extended (low IPR) and localized (high IPR) modes by the energy-dependent SPME.
In the following, we study the quench dynamics governed by this Hamiltonian to explore the consequences of this mixed spectral structure.


\textit{Entanglement Dynamics.---}
We consider unitary evolution \(|\Psi(t)\rangle = e^{-iHt} |\Psi(0)\rangle\) after a quantum quench from an unentangled product state at half filling (\(N = L/2\)). We use the N\'eel state (\(|1010\dots\rangle\)) as the initial state throughout the main text. Results from different initial states are presented in the Supplemental Material (SM)~\cite{sm}. To measure the dynamics, we compute the half-chain EE, \(S(t) = -\mathrm{Tr}[\rho_{L/2}(t) \ln \rho_{L/2}(t)]\), where \(\rho_{L/2}(t)\) is the reduced density matrix of the system's left half. 

\begin{figure}[t]
    \centering
    \includegraphics[width=0.9\linewidth]{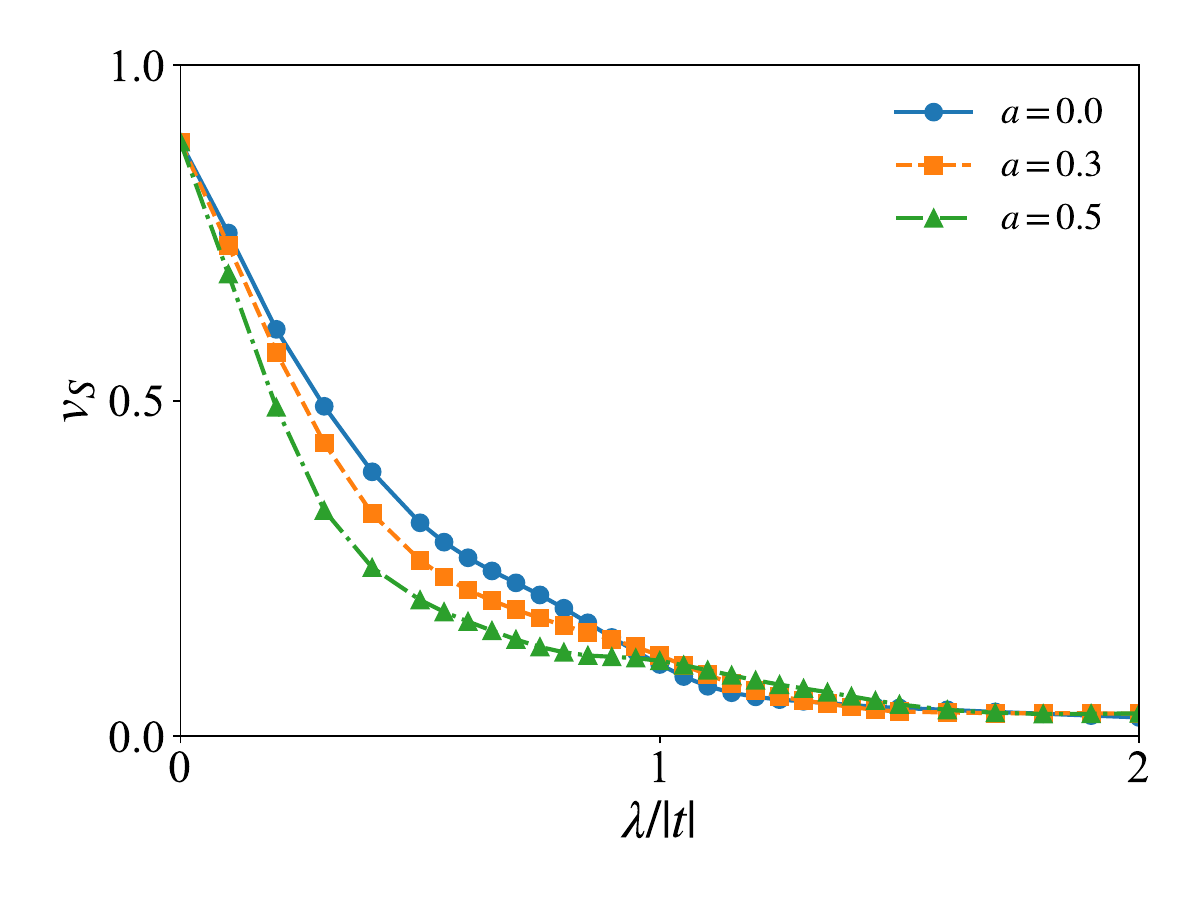}
    \caption{Early-time EE growth velocity \(v_S\) versus \(\lambda/|t|\) for different \(a\). The initial state is a N\'eel state with \(L=200\). In contrast to the saturation entanglement, \(v_S\) shows a smooth, monotonic decrease for both the AA (\(a=0\)) and GAA (\(a>0\)) models. This indicates that early-time dynamics are primarily sensitive to the potential strength rather than the detailed gap structure of the spectrum.}
    \label{fig3}
\end{figure}

First, we examine the early-time EE growth velocity, \(v_S\), extracted from the initial linear growth regime (see SM~\cite{sm} for dynamics plots and extraction details). As shown in Fig.~\ref{fig3}, for both the standard AA model (\(a=0\)) and the GAA model (\(a>0\)), \(v_S\) decreases monotonically with increasing \(\lambda\). Unlike the late-time saturation behavior discussed later, \(v_S\) does not exhibit visible plateaus corresponding to the spectral gaps. This observation highlights a fundamental \textit{dynamical dichotomy} in the system, wherein the early-time growth is governed by local kinetic constraints imposed by the potential strength \(\lambda\), rather than the global spectral topology. Consequently, the distinct gaps in the SPME spectrum have a negligible effect on the initial spreading speed. We also note that \(v_S\) begins to decrease immediately as \(\lambda > 0\), even within the fully extended phase. This confirms that any quasiperiodic potential, regardless of strength, suppresses particle propagation speed, consistent with findings in the standard AA model~\cite{longhi2021phase, shimasaki2024reversible}.

\begin{figure}
    \centering
\includegraphics[width=0.9\linewidth]{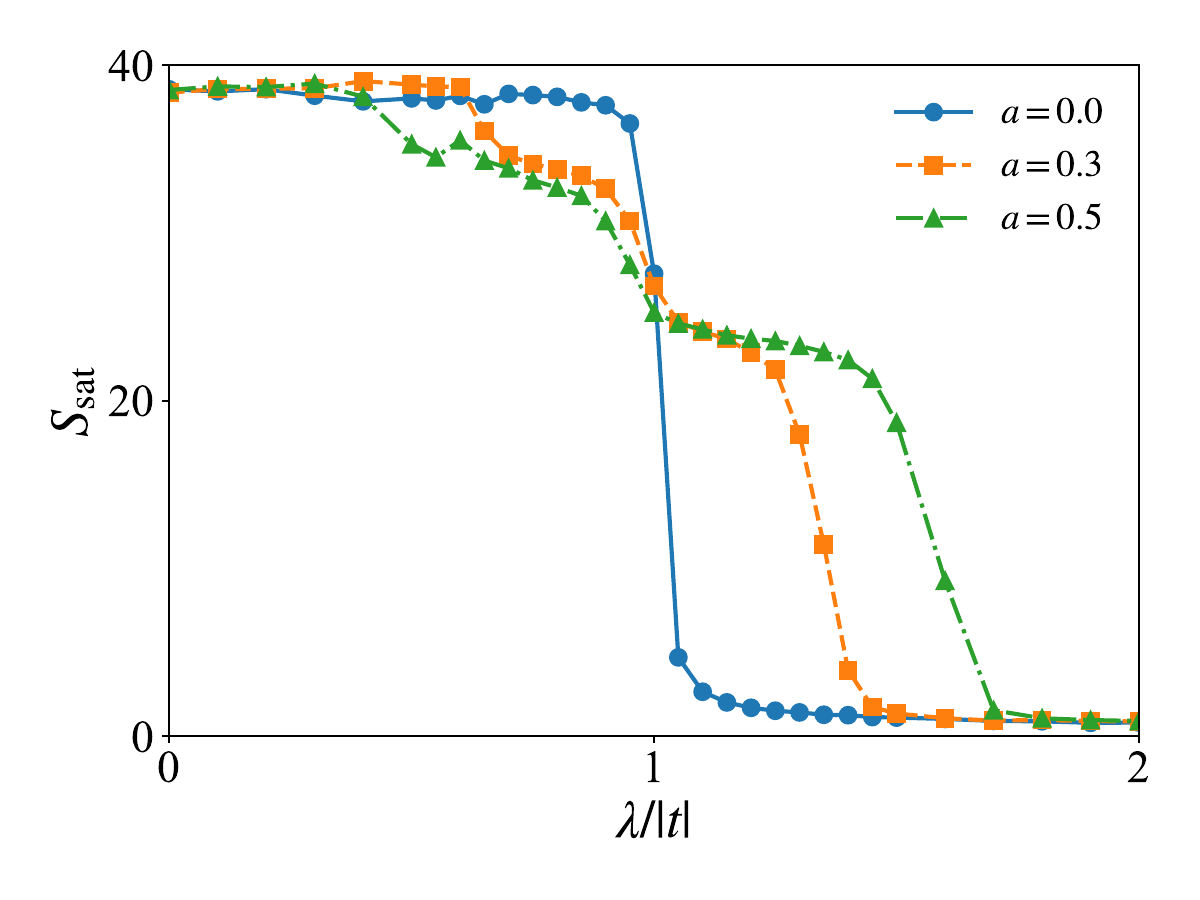}
    \caption{Saturation EE \(S_{\mathrm{sat}}\) versus \(\lambda/|t|\) after a quench from the N\'eel state for \(L = 200\). The AA model (\(a = 0\), blue) has a sharp drop at the \(\lambda = t\) transition. The GAA models (\(a > 0\)) show a smooth crossover, indicating partial delocalization caused by the SPME.}
    \label{fig4}
\end{figure}

Next, we analyze the long-time saturation EE, \(S_{\mathrm{sat}}\), extracted by averaging the EE over a long-time window after it reaches a steady value. As shown in Fig.~\ref{fig4}, for the standard AA model (\(a=0\)), \(S_{\mathrm{sat}}\) drops sharply at the \(\lambda=t\) transition, signaling the phase transition nature from a volume-law to an area-law phase. In contrast, the GAA models (\(a>0\)) exhibit a smooth crossover, indicating a gradual suppression of entanglement generation rather than a complete halt. The visible plateaus correspond to the SPME traversing the spectral gaps.

\begin{figure}
    \centering
    \includegraphics[width=0.95\linewidth]{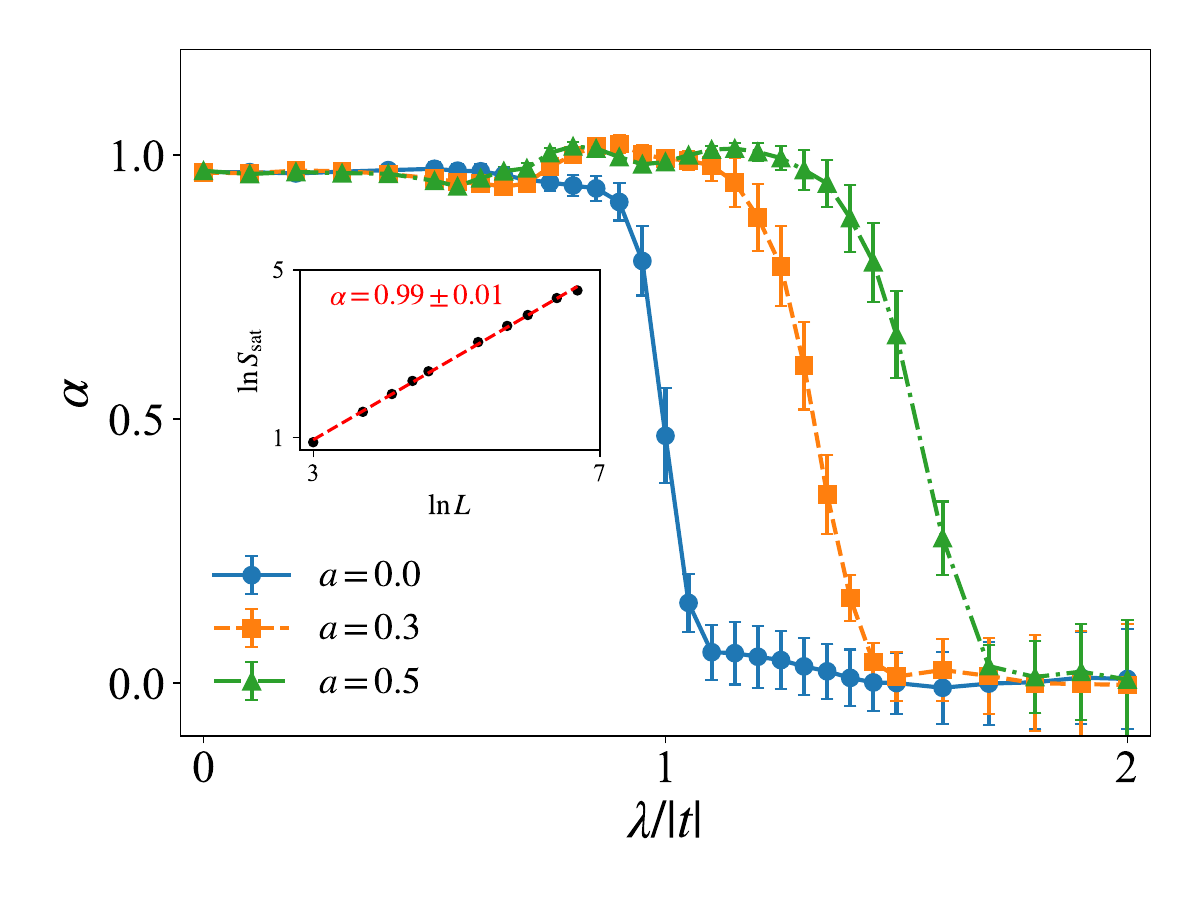}
    \caption{Scaling exponent \(\alpha\) of the saturation entropy (\(S_{\mathrm{sat}} \propto L^\alpha\)) versus \(\lambda/|t|\). The exponent distinguishes area-law (\(\alpha \approx 0\)) from volume-law (\(\alpha \approx 1\)) scaling. Error bars indicate the standard error of the slope obtained from the linear regression fit. The inset shows a representative linear fit of \(\ln S_{\mathrm{sat}}\) versus \(\ln L\) for \(a=0.3\) and \(\lambda/|t|=1.0\). The persistent volume law for \(a>0\) at intermediate \(\lambda\) confirms the delocalizing effect of the SPME.}
    \label{fig5}
\end{figure}

To rigorously probe the scaling law for the entanglement, we perform a finite-size scaling analysis of \(S_{\mathrm{sat}}\) (Fig.~\ref{fig5}) to extract the scaling exponent \(\alpha\) from the relation \(S_{\mathrm{sat}} \propto L^\alpha\), which distinguishes area-law scaling (\(\alpha \approx 0\)) from volume-law scaling (\(\alpha \approx 1\)). All cases show volume-law behavior for small \(\lambda\) and area-law for large \(\lambda\). Importantly, in the intermediate phase with SPME for \(a \neq 0\) (e.g., \(\lambda/|t|=1.0, 1.3\)), we find a persistent volume law, with the exponent \(\alpha\) remaining approximately \(1\). This confirms that the system remains partially delocalized and capable of generating extensive entanglement throughout this phase, a defining feature absent in the standard AA model where \(\lambda > t\) implies full localization and an area law.

\begin{figure}
    \centering
    \includegraphics[width=0.48\textwidth]{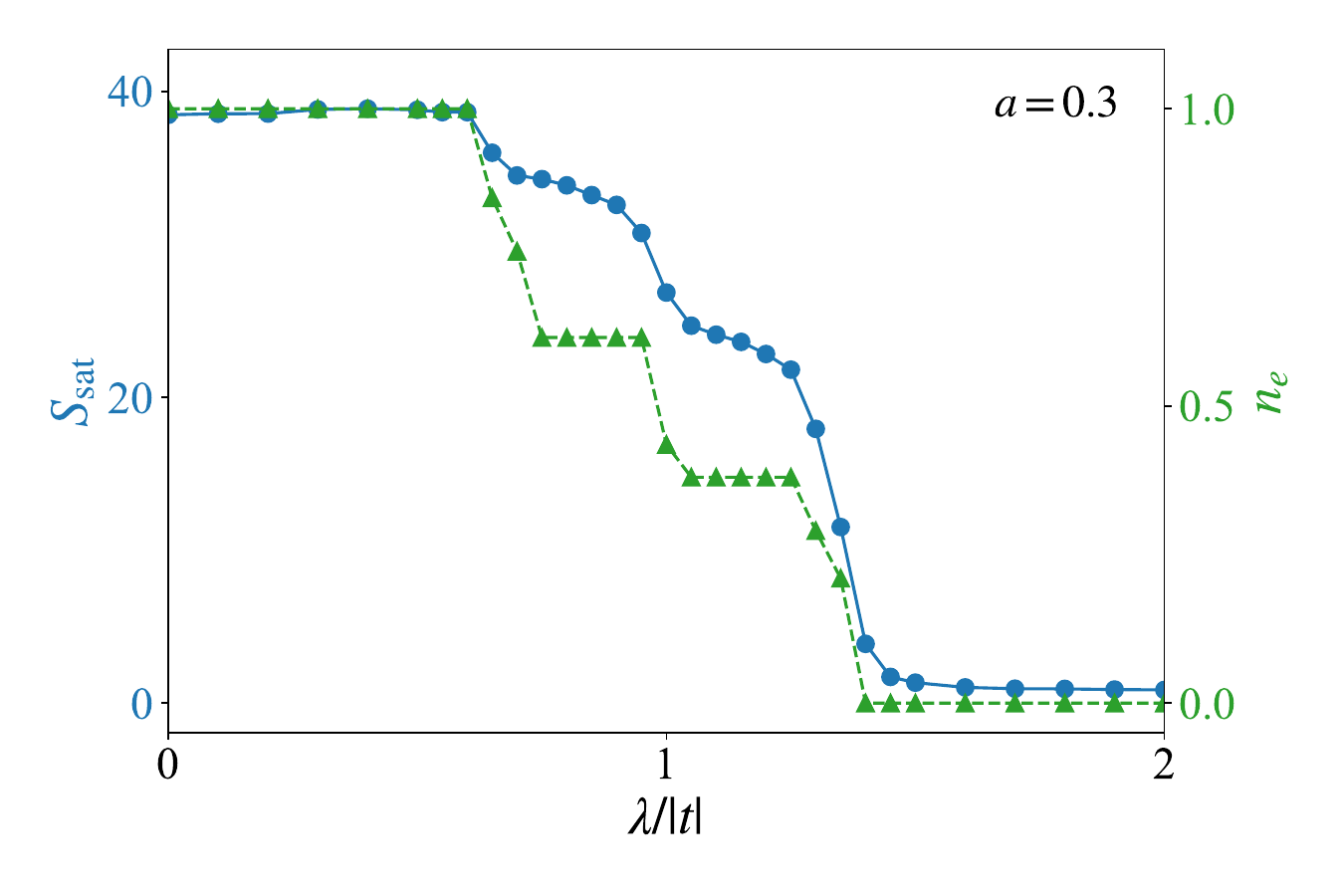} 
    \caption{Connection between saturation entanglement \(S_{\mathrm{sat}}\) (blue circles, left axis) and the fraction of extended states \(n_e\) (green triangles, right axis) versus \(\lambda/|t|\) for \(a=0.3\) and \(L=200\). The fraction is defined as \(n_e = N_e/L\), where \(N_e\) is the number of eigenstates with energy \(E < E_c\). The strong correlation shows that the system's capability to develop volume-law entanglement is directly controlled by the number of available extended modes.}
    \label{fig6}
\end{figure}

The underlying physical mechanism is revealed in Fig.~\ref{fig6}, which plots \(S_{\mathrm{sat}}\) against the fraction of extended states, \(n_e\) (defined as \(N_e/L\), where \(N_e\) counts eigenstates with \(E < E_c\)). The two quantities are strongly correlated. This striking one-to-one correspondence demonstrates that the system's capacity to generate volume-law entanglement is directly controlled by the fraction of available extended modes, which act as resources for thermalization. In other words, the decrease in \(S_{\mathrm{sat}}\) tracks the reduction in available extended modes as \(\lambda\) increases.
\begin{figure}
    \centering
    \includegraphics[width=0.42\textwidth]{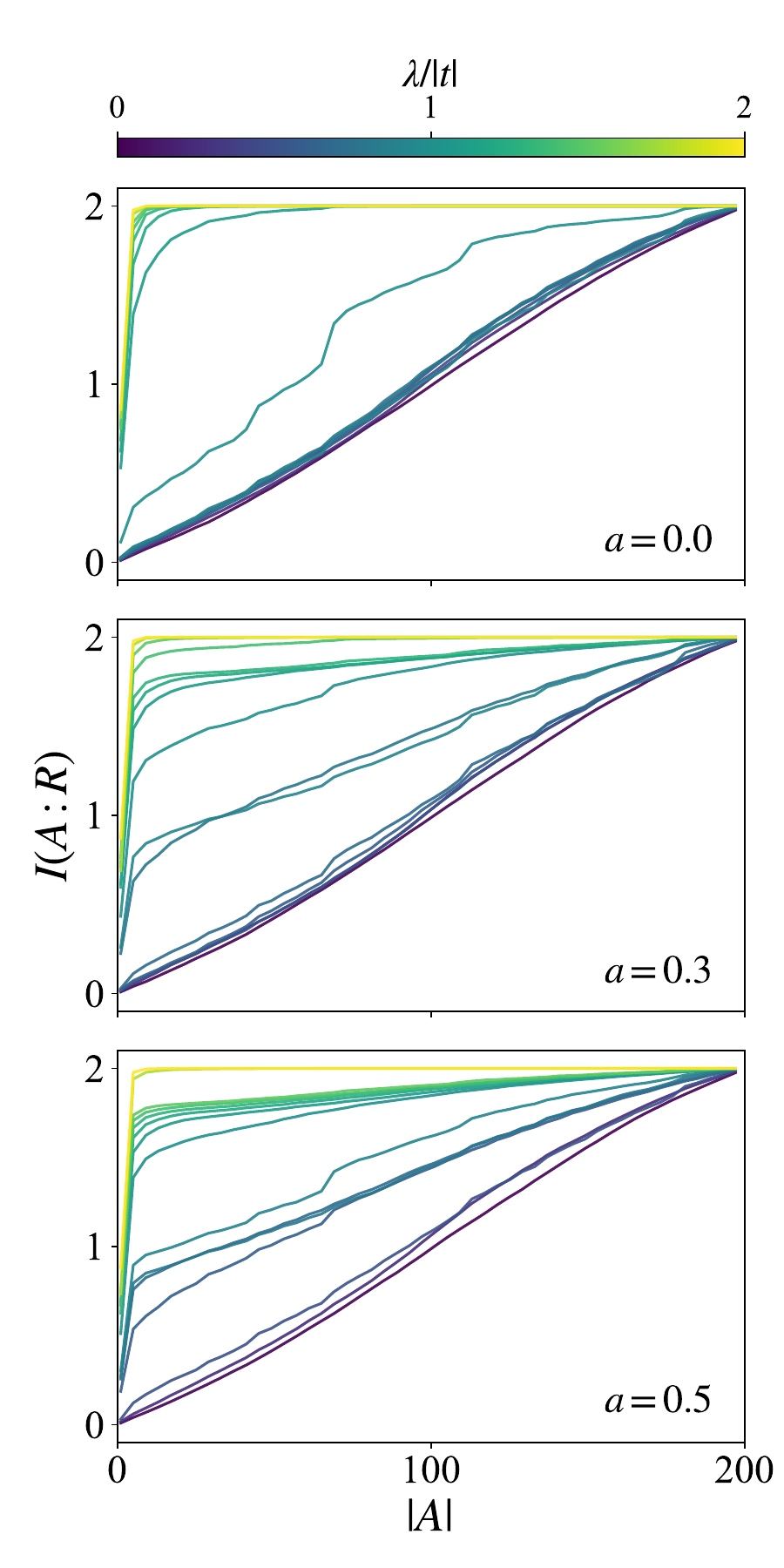} 
    \caption{Steady-state SIC profile \(I(A:R)\) versus subsystem size \(|A|\) for \(L=200\). The reference qubit is entangled with the center site \(E=L/2\). (Top) The AA model (\(a=0\)) shows a sharp transition from a linear ramp to a step function. (Middle, Bottom) The GAA model (\(a>0\)) shows a smooth crossover. The hybrid nature of the profiles in the SPME phase—an initial jump followed by a slow ramp—is a direct visualization of the coexisting localized (information trapping) and extended (information spreading) modes.}
    \label{fig7}
\end{figure}


\textit{Subsystem Information Capacity.---}
To move beyond the single scalar value of EE and gain a spatially resolved picture of information dynamics, we analyze the SIC~\cite{chen2025subsystem}. This probe measures how quantum information, initially localized at a single site, propagates through the system.  We prepare the system with a reference qubit \(R\) maximally entangled with a single site \(E\) of the chain
\begin{equation} \label{eq:bell_pair}
    |\Phi \rangle = \frac{1}{\sqrt{2}}(|1\rangle_E |0\rangle_R + |0\rangle_E |1\rangle_R),
\end{equation}
with the rest of the chain in a product state. We compute the SIC between \(R\) and a subsystem \(A\) of size \(|A|\) centered on \(E=L/2\) after the time evolution, defined as
\begin{equation} \label{eq:sic}
I(A:R) = S(A) + S(R) - S(AR),
\end{equation}
where \(S(\cdot)\) is the von Neumann entropy. Note that we use base-$2$ logarithmic for the entropy convention in the SIC definition so that $I=2$ indicates the full information recovery capability.
The spatial profile of steady-state \(I(A:R)\) measures how information spreads, with a linear ramp signifying ballistic transport in extended systems and a step-function profile indicating information trapping in localized systems~\cite{chen2025subsystem}. 

The steady-state SIC profile (Fig.~\ref{fig7}) provides a clear picture of the information distribution at late times. For the standard AA model (\(a=0\)), the profile shows a sharp transition from a perfect linear ramp (\(\lambda/|t| < 1\)) to a sharp step function (\(\lambda/|t| > 1\)). In contrast, the GAA model (\(a>0\)) shows a more smooth crossover. In the intermediate SPME regime, the SIC profiles are a hybrid of both behaviors, characterized by a steep rise for small (\(\vert A \vert\)) due to information trapping by localized modes, and a subsequent slow increase for larger (\(\vert A\vert\)) driven by ballistic spreading through the remaining extended modes. This interpolation between a linear ramp and information trapping is a direct and powerful dynamical signature of the SPME.

\begin{figure}[t]
    \centering
    \includegraphics[width=0.48\textwidth]{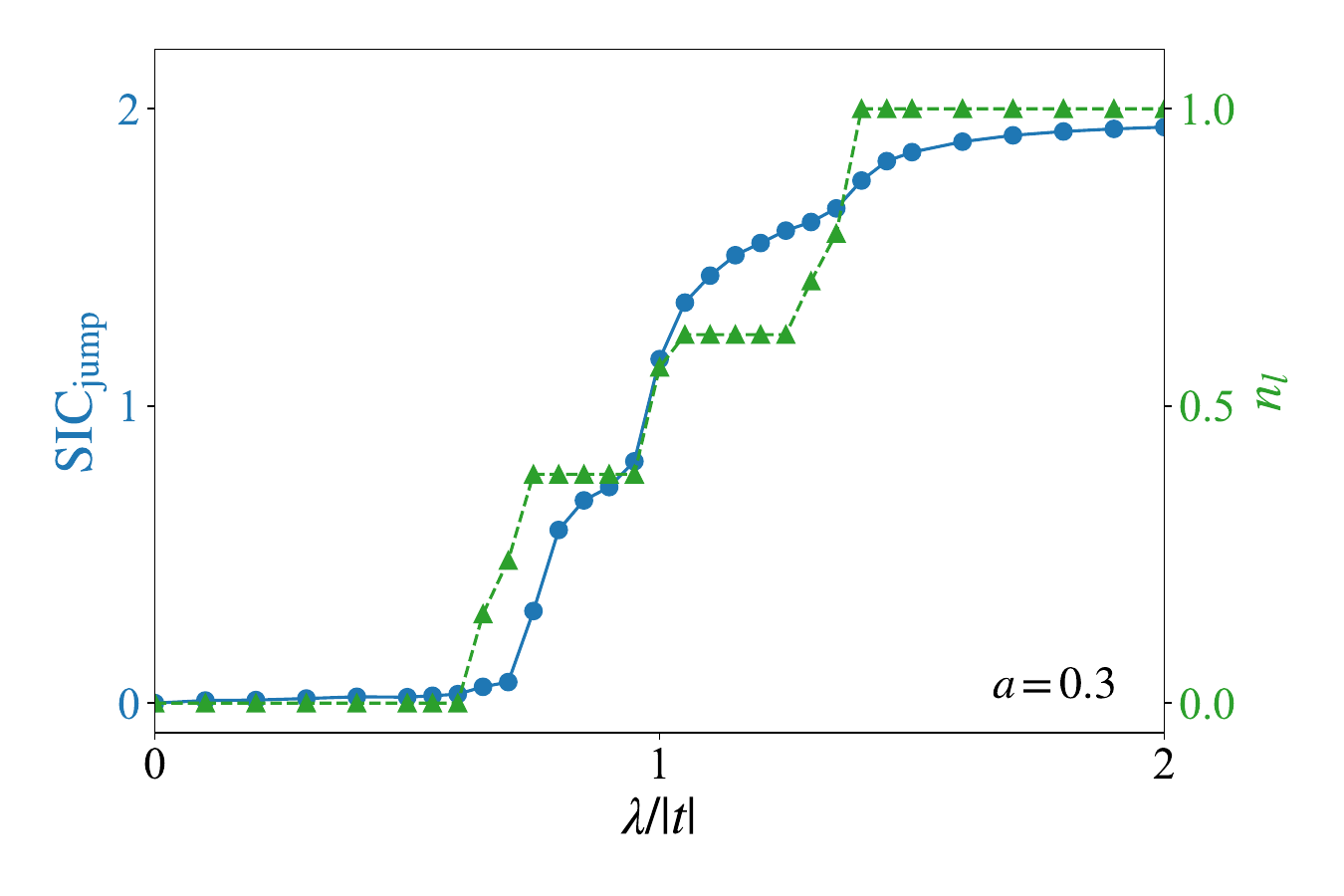}
    \caption{Connection between the initial SIC jump, \(\text{SIC}_{\text{jump}}\) (blue circles, left axis), and the fraction of localized states \(n_l\) (green triangles, right axis) for the GAA model with \(a=0.3\) and \(L=200\). The fraction is defined as \(n_l = N_l/L\), where \(N_l\) is the number of eigenstates with energy \(E > E_c\). The close connection shows that the emergence of information confinement, captured in \(\text{SIC}_{\text{jump}}\), is directly caused by the increase in localized states.}
    \label{fig8}
\end{figure}

To quantify the mixed behavior, we define \(\text{SIC}_{\text{jump}}\) as the SIC value for a small subsystem size (we use \(|A|=5\)) intended to capture the information confined by localized modes due to the trapping nature. Fig.~\ref{fig8} plots \(\text{SIC}_{\text{jump}}\) and the fraction of localized states, \(n_l\) (defined as \(N_l/L\), where \(N_l\) counts eigenstates with \(E > E_c\)). For small \(\lambda\) in the fully extended phase, \(\text{SIC}_{\text{jump}}\) is near zero. As \(\lambda\) increases in the intermediate phase with SPME, \(\text{SIC}_{\text{jump}}\) rises sharply, tracking the increase in \(n_l\). This powerful correlation between \(\text{SIC}_{\text{jump}}\) and the fraction of localized states \(n_l\) confirms the physical origin of the mixed SIC profile, revealing that the initial jump is a direct consequence of the growing fraction of localized states, which trap a corresponding amount of information around the initial site. We also present the late-time SIC profiles when the reference site is coupled to the boundary site in the SM~\cite{sm}, which further corroborates our findings.

\begin{table}[h!]
\centering
\caption{Comparison of dynamical signatures in the standard AA and GAA models within SPME phases.}
\label{tab:comparison}
\setlength{\tabcolsep}{4pt}      
\renewcommand{\arraystretch}{1.3} 

\begin{tabular*}{\columnwidth}{@{\extracolsep{\fill}} l l l}
\hline
\hline
\multirow{2}{*}{\textbf{Probe}} & \textbf{Standard AA } & \textbf{GAA } \\
&($a=0$)&($a>0$)\\
\hline
\multirow{2}{*}{\textbf{EE ($S_{\text{sat}}$)}} & Sharp transition: & SPME phase: \\
& Volume $\to$ Area & {Persistent volume-law} \\
\hline
\multirow{2}{*}{\textbf{SIC Profile}} & Sharp transition: & SPME phase: \\
& ramp $\to$ step-func. & Hybrid jump+ramp  \\

\hline
\hline
\end{tabular*}
\end{table}


\textit{Discussions and conclusion.---}Our comprehensive analysis reveals that the SPME fundamentally reshapes the system's dynamical response. Our work goes beyond the qualitative expectation of a ``smoothed'' transition to uncover the precise dynamical mechanisms governing this crossover. Our results show clear differences between the standard AA model ($a=0$) and the GAA model ($a \neq 0$) with new physical insights.
 In the AA model (\(a=0\)), the transition is sudden, with \(S_{\mathrm{sat}}\) switches from volume-law to area-law scaling abruptly, and the SIC profile changes from a ramp to a step. In contrast, the GAA model (\(a \neq 0\)) shows a smooth crossover. This is characterized by intermediate \(S_{\mathrm{sat}}\) values, mixed SIC profiles, and a persistent volume law with a decreasing entropy. This smoothing is a direct result of the SPME, which allows localized and extended modes to coexist and control the balance between information spreading and confinement. Our main findings are summarized in Table~\ref{tab:comparison}. Our results provide clear non-interacting dynamical signatures of a mobility edge, establishing an important reference for understanding thermalization in more complex quasiperiodic systems. These insights are relevant to the current debate on MBL in the presence of an SPME~\cite{li2015manybody, modak2015manybody, deng2017manybody,huang2023incommensurate}. While we focus on the paradigmatic GAA model, our findings on the dynamical signatures of coexisting states are expected to provide insights for other systems hosting mobility edges, such as those with long-range interactions or in higher dimensions \cite{li2016quantum, zhang2018universal, huang2023incommensurate}.

In conclusion, we have studied the quantum dynamics in the generalized Aubry-Andr\'e model, a paradigmatic system with a single-particle mobility edge. By leveraging the complementary power of entanglement entropy and subsystem information capacity, we have demonstrated that the SPME fundamentally reshapes the system's dynamical response, converting the sharp localization transition into a smooth crossover.  Instead of the sudden transition in the standard AA model, the SPME facilitates a regime with smoothly varying entanglement, hybrid information-spreading profiles, and a robust volume law with tunable entropy. Our findings highlight that these quantum information probes are highly sensitive to a system's spectral structure and that mobility edges reshape dynamical phase boundaries. This work provides a clear and crucial non-interacting reference for localization dynamics in quasiperiodic systems and paves the way for future studies of interacting systems, where the interplay between SPMEs and many-body effects remains a major open question. Furthermore, understanding the control over information flow offered by an SPME, from ballistic transport to perfect trapping, holds potential relevance for quantum technologies.

\textbf{Acknowledgements.} S.X.Z is supported by Quantum Science and Technology-National Science and Technology Major Project (No. 2024ZD0301700) and the National Natural Science Foundation of China (No. 12574546). Y.Q.C is supported by National Natural Science Foundation of China (No. 12504599) and NSAF (No. U2330401).

\textbf{Data availability.}  Code implementation and numerical data for this manuscript are publicly accessible in Ref. \footnote{\url{https://gitee.com/yuqiqing/sic_in_-gaa_model}}.

\let\oldaddcontentsline\addcontentsline
\renewcommand{\addcontentsline}[3]{}
\bibliography{ref}

\begin{thebibliography}{57}%
\makeatletter
\providecommand \@ifxundefined [1]{%
 \@ifx{#1\undefined}
}%
\providecommand \@ifnum [1]{%
 \ifnum #1\expandafter \@firstoftwo
 \else \expandafter \@secondoftwo
 \fi
}%
\providecommand \@ifx [1]{%
 \ifx #1\expandafter \@firstoftwo
 \else \expandafter \@secondoftwo
 \fi
}%
\providecommand \natexlab [1]{#1}%
\providecommand \enquote  [1]{``#1''}%
\providecommand \bibnamefont  [1]{#1}%
\providecommand \bibfnamefont [1]{#1}%
\providecommand \citenamefont [1]{#1}%
\providecommand \href@noop [0]{\@secondoftwo}%
\providecommand \href [0]{\begingroup \@sanitize@url \@href}%
\providecommand \@href[1]{\@@startlink{#1}\@@href}%
\providecommand \@@href[1]{\endgroup#1\@@endlink}%
\providecommand \@sanitize@url [0]{\catcode `\\12\catcode `\$12\catcode `\&12\catcode `\#12\catcode `\^12\catcode `\_12\catcode `\%12\relax}%
\providecommand \@@startlink[1]{}%
\providecommand \@@endlink[0]{}%
\providecommand \url  [0]{\begingroup\@sanitize@url \@url }%
\providecommand \@url [1]{\endgroup\@href {#1}{\urlprefix }}%
\providecommand \urlprefix  [0]{URL }%
\providecommand \Eprint [0]{\href }%
\providecommand \doibase [0]{http://dx.doi.org/}%
\providecommand \selectlanguage [0]{\@gobble}%
\providecommand \bibinfo  [0]{\@secondoftwo}%
\providecommand \bibfield  [0]{\@secondoftwo}%
\providecommand \translation [1]{[#1]}%
\providecommand \BibitemOpen [0]{}%
\providecommand \bibitemStop [0]{}%
\providecommand \bibitemNoStop [0]{.\EOS\space}%
\providecommand \EOS [0]{\spacefactor3000\relax}%
\providecommand \BibitemShut  [1]{\csname bibitem#1\endcsname}%
\let\auto@bib@innerbib\@empty
\bibitem [{\citenamefont {D'Alessio}\ \emph {et~al.}(2016)\citenamefont {D'Alessio}, \citenamefont {Kafri}, \citenamefont {Polkovnikov},\ and\ \citenamefont {Rigol}}]{dalessio2016quantum}%
  \BibitemOpen
  \bibfield  {author} {\bibinfo {author} {\bibfnamefont {Luca}\ \bibnamefont {D'Alessio}}, \bibinfo {author} {\bibfnamefont {Yariv}\ \bibnamefont {Kafri}}, \bibinfo {author} {\bibfnamefont {Anatoli}\ \bibnamefont {Polkovnikov}}, \ and\ \bibinfo {author} {\bibfnamefont {Marcos}\ \bibnamefont {Rigol}},\ }\bibfield  {title} {\enquote {\bibinfo {title} {From quantum chaos and eigenstate thermalization to statistical mechanics and thermodynamics},}\ }\href {\doibase 10.1080/00018732.2016.1198134} {\bibfield  {journal} {\bibinfo  {journal} {Advances in Physics}\ }\textbf {\bibinfo {volume} {65}},\ \bibinfo {pages} {239--362} (\bibinfo {year} {2016})}\BibitemShut {NoStop}%
\bibitem [{\citenamefont {Deutsch}(2018)}]{deutsch2018eigenstate}%
  \BibitemOpen
  \bibfield  {author} {\bibinfo {author} {\bibfnamefont {Joshua~M}\ \bibnamefont {Deutsch}},\ }\bibfield  {title} {\enquote {\bibinfo {title} {Eigenstate thermalization hypothesis},}\ }\href {\doibase 10.1088/1361-6633/aac9f1} {\bibfield  {journal} {\bibinfo  {journal} {Reports on Progress in Physics}\ }\textbf {\bibinfo {volume} {81}},\ \bibinfo {pages} {082001} (\bibinfo {year} {2018})}\BibitemShut {NoStop}%
\bibitem [{\citenamefont {Deutsch}(1991)}]{deutsch1991quantum}%
  \BibitemOpen
  \bibfield  {author} {\bibinfo {author} {\bibfnamefont {J.~M.}\ \bibnamefont {Deutsch}},\ }\bibfield  {title} {\enquote {\bibinfo {title} {Quantum statistical mechanics in a closed system},}\ }\href {\doibase 10.1103/PhysRevA.43.2046} {\bibfield  {journal} {\bibinfo  {journal} {Physical Review A}\ }\textbf {\bibinfo {volume} {43}},\ \bibinfo {pages} {2046--2049} (\bibinfo {year} {1991})}\BibitemShut {NoStop}%
\bibitem [{\citenamefont {Srednicki}(1994)}]{srednicki1994chaos}%
  \BibitemOpen
  \bibfield  {author} {\bibinfo {author} {\bibfnamefont {Mark}\ \bibnamefont {Srednicki}},\ }\bibfield  {title} {\enquote {\bibinfo {title} {Chaos and quantum thermalization},}\ }\href {\doibase 10.1103/PhysRevE.50.888} {\bibfield  {journal} {\bibinfo  {journal} {Physical Review E}\ }\textbf {\bibinfo {volume} {50}},\ \bibinfo {pages} {888--901} (\bibinfo {year} {1994})}\BibitemShut {NoStop}%
\bibitem [{\citenamefont {Rigol}\ \emph {et~al.}(2008)\citenamefont {Rigol}, \citenamefont {Dunjko},\ and\ \citenamefont {Olshanii}}]{rigol2008thermalization}%
  \BibitemOpen
  \bibfield  {author} {\bibinfo {author} {\bibfnamefont {Marcos}\ \bibnamefont {Rigol}}, \bibinfo {author} {\bibfnamefont {Vanja}\ \bibnamefont {Dunjko}}, \ and\ \bibinfo {author} {\bibfnamefont {Maxim}\ \bibnamefont {Olshanii}},\ }\bibfield  {title} {\enquote {\bibinfo {title} {Thermalization and its mechanism for generic isolated quantum systems},}\ }\href {\doibase 10.1038/nature06838} {\bibfield  {journal} {\bibinfo  {journal} {Nature}\ }\textbf {\bibinfo {volume} {452}},\ \bibinfo {pages} {854--858} (\bibinfo {year} {2008})}\BibitemShut {NoStop}%
\bibitem [{\citenamefont {Anderson}(1958)}]{anderson1958absence}%
  \BibitemOpen
  \bibfield  {author} {\bibinfo {author} {\bibfnamefont {P.~W.}\ \bibnamefont {Anderson}},\ }\bibfield  {title} {\enquote {\bibinfo {title} {Absence of diffusion in certain random lattices},}\ }\href {\doibase 10.1103/PhysRev.109.1492} {\bibfield  {journal} {\bibinfo  {journal} {Physical Review}\ }\textbf {\bibinfo {volume} {109}},\ \bibinfo {pages} {1492--1505} (\bibinfo {year} {1958})}\BibitemShut {NoStop}%
\bibitem [{\citenamefont {Fleishman}\ and\ \citenamefont {Anderson}(1980)}]{fleishman1980interactions}%
  \BibitemOpen
  \bibfield  {author} {\bibinfo {author} {\bibfnamefont {L.}~\bibnamefont {Fleishman}}\ and\ \bibinfo {author} {\bibfnamefont {P.~W.}\ \bibnamefont {Anderson}},\ }\bibfield  {title} {\enquote {\bibinfo {title} {Interactions and the anderson transition},}\ }\href {\doibase 10.1103/PhysRevB.21.2366} {\bibfield  {journal} {\bibinfo  {journal} {Physical Review B}\ }\textbf {\bibinfo {volume} {21}},\ \bibinfo {pages} {2366--2377} (\bibinfo {year} {1980})}\BibitemShut {NoStop}%
\bibitem [{\citenamefont {Basko}\ \emph {et~al.}(2006)\citenamefont {Basko}, \citenamefont {Aleiner},\ and\ \citenamefont {Altshuler}}]{basko2006metal}%
  \BibitemOpen
  \bibfield  {author} {\bibinfo {author} {\bibfnamefont {D.~M.}\ \bibnamefont {Basko}}, \bibinfo {author} {\bibfnamefont {I.~L.}\ \bibnamefont {Aleiner}}, \ and\ \bibinfo {author} {\bibfnamefont {B.~L.}\ \bibnamefont {Altshuler}},\ }\bibfield  {title} {\enquote {\bibinfo {title} {Metal--insulator transition in a weakly interacting many-electron system with localized single-particle states},}\ }\href {\doibase 10.1016/j.aop.2005.11.014} {\bibfield  {journal} {\bibinfo  {journal} {Annals of Physics}\ }\textbf {\bibinfo {volume} {321}},\ \bibinfo {pages} {1126--1205} (\bibinfo {year} {2006})}\BibitemShut {NoStop}%
\bibitem [{\citenamefont {Nandkishore}\ and\ \citenamefont {Huse}(2015)}]{nandkishore2015manybody}%
  \BibitemOpen
  \bibfield  {author} {\bibinfo {author} {\bibfnamefont {Rahul}\ \bibnamefont {Nandkishore}}\ and\ \bibinfo {author} {\bibfnamefont {David~A.}\ \bibnamefont {Huse}},\ }\bibfield  {title} {\enquote {\bibinfo {title} {Many-body localization and thermalization in quantum statistical mechanics},}\ }\href {\doibase 10.1146/annurev-conmatphys-031214-014726} {\bibfield  {journal} {\bibinfo  {journal} {Annual Review of Condensed Matter Physics}\ }\textbf {\bibinfo {volume} {6}},\ \bibinfo {pages} {15--38} (\bibinfo {year} {2015})}\BibitemShut {NoStop}%
\bibitem [{\citenamefont {Abanin}\ \emph {et~al.}(2019)\citenamefont {Abanin}, \citenamefont {Altman}, \citenamefont {Bloch},\ and\ \citenamefont {Serbyn}}]{abanin2019colloquium}%
  \BibitemOpen
  \bibfield  {author} {\bibinfo {author} {\bibfnamefont {Dmitry~A.}\ \bibnamefont {Abanin}}, \bibinfo {author} {\bibfnamefont {Ehud}\ \bibnamefont {Altman}}, \bibinfo {author} {\bibfnamefont {Immanuel}\ \bibnamefont {Bloch}}, \ and\ \bibinfo {author} {\bibfnamefont {Maksym}\ \bibnamefont {Serbyn}},\ }\bibfield  {title} {\enquote {\bibinfo {title} {Colloquium: Many-body localization, thermalization, and entanglement},}\ }\href {\doibase 10.1103/RevModPhys.91.021001} {\bibfield  {journal} {\bibinfo  {journal} {Reviews of Modern Physics}\ }\textbf {\bibinfo {volume} {91}},\ \bibinfo {pages} {021001} (\bibinfo {year} {2019})}\BibitemShut {NoStop}%
\bibitem [{\citenamefont {Thouless}(1983)}]{thouless1983quantization}%
  \BibitemOpen
  \bibfield  {author} {\bibinfo {author} {\bibfnamefont {D.~J.}\ \bibnamefont {Thouless}},\ }\bibfield  {title} {\enquote {\bibinfo {title} {Quantization of particle transport},}\ }\href {\doibase 10.1103/PhysRevB.27.6083} {\bibfield  {journal} {\bibinfo  {journal} {Physical Review B}\ }\textbf {\bibinfo {volume} {27}},\ \bibinfo {pages} {6083--6087} (\bibinfo {year} {1983})}\BibitemShut {NoStop}%
\bibitem [{\citenamefont {Han}\ \emph {et~al.}(1994)\citenamefont {Han}, \citenamefont {Thouless}, \citenamefont {Hiramoto},\ and\ \citenamefont {Kohmoto}}]{han1994critical}%
  \BibitemOpen
  \bibfield  {author} {\bibinfo {author} {\bibfnamefont {J.~H.}\ \bibnamefont {Han}}, \bibinfo {author} {\bibfnamefont {D.~J.}\ \bibnamefont {Thouless}}, \bibinfo {author} {\bibfnamefont {H.}~\bibnamefont {Hiramoto}}, \ and\ \bibinfo {author} {\bibfnamefont {M.}~\bibnamefont {Kohmoto}},\ }\bibfield  {title} {\enquote {\bibinfo {title} {Critical and bicritical properties of harper's equation with next-nearest-neighbor coupling},}\ }\href {\doibase 10.1103/PhysRevB.50.11365} {\bibfield  {journal} {\bibinfo  {journal} {Physical Review B}\ }\textbf {\bibinfo {volume} {50}},\ \bibinfo {pages} {11365--11380} (\bibinfo {year} {1994})}\BibitemShut {NoStop}%
\bibitem [{\citenamefont {Iyer}\ \emph {et~al.}(2013)\citenamefont {Iyer}, \citenamefont {Oganesyan}, \citenamefont {Refael},\ and\ \citenamefont {Huse}}]{iyer2013manybody}%
  \BibitemOpen
  \bibfield  {author} {\bibinfo {author} {\bibfnamefont {Shankar}\ \bibnamefont {Iyer}}, \bibinfo {author} {\bibfnamefont {Vadim}\ \bibnamefont {Oganesyan}}, \bibinfo {author} {\bibfnamefont {Gil}\ \bibnamefont {Refael}}, \ and\ \bibinfo {author} {\bibfnamefont {David~A.}\ \bibnamefont {Huse}},\ }\bibfield  {title} {\enquote {\bibinfo {title} {Many-body localization in a quasiperiodic system},}\ }\href {\doibase 10.1103/PhysRevB.87.134202} {\bibfield  {journal} {\bibinfo  {journal} {Physical Review B}\ }\textbf {\bibinfo {volume} {87}},\ \bibinfo {pages} {134202} (\bibinfo {year} {2013})}\BibitemShut {NoStop}%
\bibitem [{\citenamefont {Schreiber}\ \emph {et~al.}(2015)\citenamefont {Schreiber}, \citenamefont {Hodgman}, \citenamefont {Bordia}, \citenamefont {L{\"u}schen}, \citenamefont {Fischer}, \citenamefont {Vosk}, \citenamefont {Altman}, \citenamefont {Schneider},\ and\ \citenamefont {Bloch}}]{schreiber2015observation}%
  \BibitemOpen
  \bibfield  {author} {\bibinfo {author} {\bibfnamefont {Michael}\ \bibnamefont {Schreiber}}, \bibinfo {author} {\bibfnamefont {Sean~S.}\ \bibnamefont {Hodgman}}, \bibinfo {author} {\bibfnamefont {Pranjal}\ \bibnamefont {Bordia}}, \bibinfo {author} {\bibfnamefont {Henrik~P.}\ \bibnamefont {L{\"u}schen}}, \bibinfo {author} {\bibfnamefont {Mark~H.}\ \bibnamefont {Fischer}}, \bibinfo {author} {\bibfnamefont {Ronen}\ \bibnamefont {Vosk}}, \bibinfo {author} {\bibfnamefont {Ehud}\ \bibnamefont {Altman}}, \bibinfo {author} {\bibfnamefont {Ulrich}\ \bibnamefont {Schneider}}, \ and\ \bibinfo {author} {\bibfnamefont {Immanuel}\ \bibnamefont {Bloch}},\ }\bibfield  {title} {\enquote {\bibinfo {title} {Observation of many-body localization of interacting fermions in a quasirandom optical lattice},}\ }\href {\doibase 10.1126/science.aaa7432} {\bibfield  {journal} {\bibinfo  {journal} {Science}\ }\textbf {\bibinfo {volume} {349}},\ \bibinfo {pages} {842--845} (\bibinfo {year} {2015})}\BibitemShut {NoStop}%
\bibitem [{\citenamefont {Lee}\ \emph {et~al.}(2017)\citenamefont {Lee}, \citenamefont {Look}, \citenamefont {Lim},\ and\ \citenamefont {Sheng}}]{lee2017manybody}%
  \BibitemOpen
  \bibfield  {author} {\bibinfo {author} {\bibfnamefont {Mac}\ \bibnamefont {Lee}}, \bibinfo {author} {\bibfnamefont {Thomas~R.}\ \bibnamefont {Look}}, \bibinfo {author} {\bibfnamefont {S.~P.}\ \bibnamefont {Lim}}, \ and\ \bibinfo {author} {\bibfnamefont {D.~N.}\ \bibnamefont {Sheng}},\ }\bibfield  {title} {\enquote {\bibinfo {title} {Many-body localization in spin chain systems with quasiperiodic fields},}\ }\href {\doibase 10.1103/PhysRevB.96.075146} {\bibfield  {journal} {\bibinfo  {journal} {Physical Review B}\ }\textbf {\bibinfo {volume} {96}},\ \bibinfo {pages} {075146} (\bibinfo {year} {2017})}\BibitemShut {NoStop}%
\bibitem [{\citenamefont {Khemani}\ \emph {et~al.}(2017)\citenamefont {Khemani}, \citenamefont {Sheng},\ and\ \citenamefont {Huse}}]{khemani2017two}%
  \BibitemOpen
  \bibfield  {author} {\bibinfo {author} {\bibfnamefont {Vedika}\ \bibnamefont {Khemani}}, \bibinfo {author} {\bibfnamefont {D.~N.}\ \bibnamefont {Sheng}}, \ and\ \bibinfo {author} {\bibfnamefont {David~A.}\ \bibnamefont {Huse}},\ }\bibfield  {title} {\enquote {\bibinfo {title} {Two universality classes for the many-body localization transition},}\ }\href {\doibase 10.1103/PhysRevLett.119.075702} {\bibfield  {journal} {\bibinfo  {journal} {Physical Review Letters}\ }\textbf {\bibinfo {volume} {119}},\ \bibinfo {pages} {075702} (\bibinfo {year} {2017})}\BibitemShut {NoStop}%
\bibitem [{\citenamefont {Chandran}\ and\ \citenamefont {Laumann}(2017)}]{chandran2017localization}%
  \BibitemOpen
  \bibfield  {author} {\bibinfo {author} {\bibfnamefont {A.}~\bibnamefont {Chandran}}\ and\ \bibinfo {author} {\bibfnamefont {C.~R.}\ \bibnamefont {Laumann}},\ }\bibfield  {title} {\enquote {\bibinfo {title} {Localization and symmetry breaking in the quantum quasiperiodic ising glass},}\ }\href {\doibase 10.1103/PhysRevX.7.031061} {\bibfield  {journal} {\bibinfo  {journal} {Physical Review X}\ }\textbf {\bibinfo {volume} {7}},\ \bibinfo {pages} {031061} (\bibinfo {year} {2017})}\BibitemShut {NoStop}%
\bibitem [{\citenamefont {Setiawan}\ \emph {et~al.}(2017)\citenamefont {Setiawan}, \citenamefont {Deng},\ and\ \citenamefont {Pixley}}]{setiawan2017transport}%
  \BibitemOpen
  \bibfield  {author} {\bibinfo {author} {\bibfnamefont {F.}~\bibnamefont {Setiawan}}, \bibinfo {author} {\bibfnamefont {Dong-Ling}\ \bibnamefont {Deng}}, \ and\ \bibinfo {author} {\bibfnamefont {J.~H.}\ \bibnamefont {Pixley}},\ }\bibfield  {title} {\enquote {\bibinfo {title} {Transport properties across the many-body localization transition in quasiperiodic and random systems},}\ }\href {\doibase 10.1103/PhysRevB.96.104205} {\bibfield  {journal} {\bibinfo  {journal} {Physical Review B}\ }\textbf {\bibinfo {volume} {96}},\ \bibinfo {pages} {104205} (\bibinfo {year} {2017})}\BibitemShut {NoStop}%
\bibitem [{\citenamefont {Zhang}\ and\ \citenamefont {Yao}(2018)}]{zhang2018universal}%
  \BibitemOpen
  \bibfield  {author} {\bibinfo {author} {\bibfnamefont {Shi-Xin}\ \bibnamefont {Zhang}}\ and\ \bibinfo {author} {\bibfnamefont {Hong}\ \bibnamefont {Yao}},\ }\bibfield  {title} {\enquote {\bibinfo {title} {Universal properties of many-body localization transitions in quasiperiodic systems},}\ }\href {\doibase 10.1103/PhysRevLett.121.206601} {\bibfield  {journal} {\bibinfo  {journal} {Physical Review Letters}\ }\textbf {\bibinfo {volume} {121}},\ \bibinfo {pages} {206601} (\bibinfo {year} {2018})}\BibitemShut {NoStop}%
\bibitem [{\citenamefont {Zhang}\ and\ \citenamefont {Yao}(2019)}]{zhang2019strong}%
  \BibitemOpen
  \bibfield  {author} {\bibinfo {author} {\bibfnamefont {Shi-Xin}\ \bibnamefont {Zhang}}\ and\ \bibinfo {author} {\bibfnamefont {Hong}\ \bibnamefont {Yao}},\ }\href {\doibase 10.48550/arXiv.1906.00971} {\enquote {\bibinfo {title} {Strong and weak many-body localizations},}\ } (\bibinfo {year} {2019}),\ \Eprint {http://arxiv.org/abs/1906.00971} {arXiv:1906.00971 [cond-mat]} \BibitemShut {NoStop}%
\bibitem [{\citenamefont {Wang}\ \emph {et~al.}(2020)\citenamefont {Wang}, \citenamefont {Xia}, \citenamefont {Zhang}, \citenamefont {Yao}, \citenamefont {Chen}, \citenamefont {You}, \citenamefont {Zhou},\ and\ \citenamefont {Liu}}]{wang2020onedimensional}%
  \BibitemOpen
  \bibfield  {author} {\bibinfo {author} {\bibfnamefont {Yucheng}\ \bibnamefont {Wang}}, \bibinfo {author} {\bibfnamefont {Xu}~\bibnamefont {Xia}}, \bibinfo {author} {\bibfnamefont {Long}\ \bibnamefont {Zhang}}, \bibinfo {author} {\bibfnamefont {Hepeng}\ \bibnamefont {Yao}}, \bibinfo {author} {\bibfnamefont {Shu}\ \bibnamefont {Chen}}, \bibinfo {author} {\bibfnamefont {Jiangong}\ \bibnamefont {You}}, \bibinfo {author} {\bibfnamefont {Qi}~\bibnamefont {Zhou}}, \ and\ \bibinfo {author} {\bibfnamefont {Xiong-Jun}\ \bibnamefont {Liu}},\ }\bibfield  {title} {\enquote {\bibinfo {title} {One-dimensional quasiperiodic mosaic lattice with exact mobility edges},}\ }\href {\doibase 10.1103/PhysRevLett.125.196604} {\bibfield  {journal} {\bibinfo  {journal} {Physical Review Letters}\ }\textbf {\bibinfo {volume} {125}},\ \bibinfo {pages} {196604} (\bibinfo {year} {2020})}\BibitemShut {NoStop}%
\bibitem [{\citenamefont {Zhai}\ \emph {et~al.}(2020)\citenamefont {Zhai}, \citenamefont {Yin},\ and\ \citenamefont {Huang}}]{zhai2020manybody}%
  \BibitemOpen
  \bibfield  {author} {\bibinfo {author} {\bibfnamefont {Liang-Jun}\ \bibnamefont {Zhai}}, \bibinfo {author} {\bibfnamefont {Shuai}\ \bibnamefont {Yin}}, \ and\ \bibinfo {author} {\bibfnamefont {Guang-Yao}\ \bibnamefont {Huang}},\ }\bibfield  {title} {\enquote {\bibinfo {title} {Many-body localization in a non-hermitian quasiperiodic system},}\ }\href {\doibase 10.1103/PhysRevB.102.064206} {\bibfield  {journal} {\bibinfo  {journal} {Physical Review B}\ }\textbf {\bibinfo {volume} {102}},\ \bibinfo {pages} {064206} (\bibinfo {year} {2020})}\BibitemShut {NoStop}%
\bibitem [{\citenamefont {Aubry}\ and\ \citenamefont {Andr{\'e}}(1980)}]{aubry1980analyticity}%
  \BibitemOpen
  \bibfield  {author} {\bibinfo {author} {\bibfnamefont {Serge}\ \bibnamefont {Aubry}}\ and\ \bibinfo {author} {\bibfnamefont {Gilles}\ \bibnamefont {Andr{\'e}}},\ }\bibfield  {title} {\enquote {\bibinfo {title} {Analyticity breaking and anderson localization in incommensurate lattices},}\ }\href@noop {} {\bibfield  {journal} {\bibinfo  {journal} {Ann. Israel Phys. Soc}\ }\textbf {\bibinfo {volume} {3}},\ \bibinfo {pages} {133} (\bibinfo {year} {1980})}\BibitemShut {NoStop}%
\bibitem [{\citenamefont {Harper}(1955)}]{harper1955single}%
  \BibitemOpen
  \bibfield  {author} {\bibinfo {author} {\bibfnamefont {P.~G.}\ \bibnamefont {Harper}},\ }\bibfield  {title} {\enquote {\bibinfo {title} {Single band motion of conduction electrons in a uniform magnetic field},}\ }\href {\doibase 10.1088/0370-1298/68/10/304} {\bibfield  {journal} {\bibinfo  {journal} {Proceedings of the Physical Society. Section A}\ }\textbf {\bibinfo {volume} {68}},\ \bibinfo {pages} {874} (\bibinfo {year} {1955})}\BibitemShut {NoStop}%
\bibitem [{\citenamefont {Biddle}\ and\ \citenamefont {Das~Sarma}(2010)}]{biddle2010predicted}%
  \BibitemOpen
  \bibfield  {author} {\bibinfo {author} {\bibfnamefont {J.}~\bibnamefont {Biddle}}\ and\ \bibinfo {author} {\bibfnamefont {S.}~\bibnamefont {Das~Sarma}},\ }\bibfield  {title} {\enquote {\bibinfo {title} {Predicted mobility edges in one-dimensional incommensurate optical lattices: An exactly solvable model of anderson localization},}\ }\href {\doibase 10.1103/PhysRevLett.104.070601} {\bibfield  {journal} {\bibinfo  {journal} {Physical Review Letters}\ }\textbf {\bibinfo {volume} {104}},\ \bibinfo {pages} {070601} (\bibinfo {year} {2010})}\BibitemShut {NoStop}%
\bibitem [{\citenamefont {Ganeshan}\ \emph {et~al.}(2015)\citenamefont {Ganeshan}, \citenamefont {Pixley},\ and\ \citenamefont {Das~Sarma}}]{ganeshan2015nearest}%
  \BibitemOpen
  \bibfield  {author} {\bibinfo {author} {\bibfnamefont {Sriram}\ \bibnamefont {Ganeshan}}, \bibinfo {author} {\bibfnamefont {J.~H.}\ \bibnamefont {Pixley}}, \ and\ \bibinfo {author} {\bibfnamefont {S.}~\bibnamefont {Das~Sarma}},\ }\bibfield  {title} {\enquote {\bibinfo {title} {Nearest neighbor tight binding models with an exact mobility edge in one dimension},}\ }\href {\doibase 10.1103/PhysRevLett.114.146601} {\bibfield  {journal} {\bibinfo  {journal} {Physical Review Letters}\ }\textbf {\bibinfo {volume} {114}},\ \bibinfo {pages} {146601} (\bibinfo {year} {2015})}\BibitemShut {NoStop}%
\bibitem [{\citenamefont {L{\"u}schen}\ \emph {et~al.}(2018)\citenamefont {L{\"u}schen}, \citenamefont {Scherg}, \citenamefont {Kohlert}, \citenamefont {Schreiber}, \citenamefont {Bordia}, \citenamefont {Li}, \citenamefont {Das~Sarma},\ and\ \citenamefont {Bloch}}]{luschen2018singleparticle}%
  \BibitemOpen
  \bibfield  {author} {\bibinfo {author} {\bibfnamefont {Henrik~P.}\ \bibnamefont {L{\"u}schen}}, \bibinfo {author} {\bibfnamefont {Sebastian}\ \bibnamefont {Scherg}}, \bibinfo {author} {\bibfnamefont {Thomas}\ \bibnamefont {Kohlert}}, \bibinfo {author} {\bibfnamefont {Michael}\ \bibnamefont {Schreiber}}, \bibinfo {author} {\bibfnamefont {Pranjal}\ \bibnamefont {Bordia}}, \bibinfo {author} {\bibfnamefont {Xiao}\ \bibnamefont {Li}}, \bibinfo {author} {\bibfnamefont {S.}~\bibnamefont {Das~Sarma}}, \ and\ \bibinfo {author} {\bibfnamefont {Immanuel}\ \bibnamefont {Bloch}},\ }\bibfield  {title} {\enquote {\bibinfo {title} {Single-particle mobility edge in a one-dimensional quasiperiodic optical lattice},}\ }\href {\doibase 10.1103/PhysRevLett.120.160404} {\bibfield  {journal} {\bibinfo  {journal} {Physical Review Letters}\ }\textbf {\bibinfo {volume} {120}},\ \bibinfo {pages} {160404} (\bibinfo {year} {2018})}\BibitemShut {NoStop}%
\bibitem [{\citenamefont {Kohlert}\ \emph {et~al.}(2019)\citenamefont {Kohlert}, \citenamefont {Scherg}, \citenamefont {Li}, \citenamefont {L{\"u}schen}, \citenamefont {Das~Sarma}, \citenamefont {Bloch},\ and\ \citenamefont {Aidelsburger}}]{kohlert2019observation}%
  \BibitemOpen
  \bibfield  {author} {\bibinfo {author} {\bibfnamefont {Thomas}\ \bibnamefont {Kohlert}}, \bibinfo {author} {\bibfnamefont {Sebastian}\ \bibnamefont {Scherg}}, \bibinfo {author} {\bibfnamefont {Xiao}\ \bibnamefont {Li}}, \bibinfo {author} {\bibfnamefont {Henrik~P.}\ \bibnamefont {L{\"u}schen}}, \bibinfo {author} {\bibfnamefont {Sankar}\ \bibnamefont {Das~Sarma}}, \bibinfo {author} {\bibfnamefont {Immanuel}\ \bibnamefont {Bloch}}, \ and\ \bibinfo {author} {\bibfnamefont {Monika}\ \bibnamefont {Aidelsburger}},\ }\bibfield  {title} {\enquote {\bibinfo {title} {Observation of many-body localization in a one-dimensional system with a single-particle mobility edge},}\ }\href {\doibase 10.1103/PhysRevLett.122.170403} {\bibfield  {journal} {\bibinfo  {journal} {Physical Review Letters}\ }\textbf {\bibinfo {volume} {122}},\ \bibinfo {pages} {170403} (\bibinfo {year} {2019})}\BibitemShut {NoStop}%
\bibitem [{\citenamefont {Verbin}\ \emph {et~al.}(2015)\citenamefont {Verbin}, \citenamefont {Zilberberg}, \citenamefont {Lahini}, \citenamefont {Kraus},\ and\ \citenamefont {Silberberg}}]{verbin2015topological}%
  \BibitemOpen
  \bibfield  {author} {\bibinfo {author} {\bibfnamefont {Mor}\ \bibnamefont {Verbin}}, \bibinfo {author} {\bibfnamefont {Oded}\ \bibnamefont {Zilberberg}}, \bibinfo {author} {\bibfnamefont {Yoav}\ \bibnamefont {Lahini}}, \bibinfo {author} {\bibfnamefont {Yaacov~E.}\ \bibnamefont {Kraus}}, \ and\ \bibinfo {author} {\bibfnamefont {Yaron}\ \bibnamefont {Silberberg}},\ }\bibfield  {title} {\enquote {\bibinfo {title} {Topological pumping over a photonic fibonacci quasicrystal},}\ }\href {\doibase 10.1103/PhysRevB.91.064201} {\bibfield  {journal} {\bibinfo  {journal} {Physical Review B}\ }\textbf {\bibinfo {volume} {91}},\ \bibinfo {pages} {064201} (\bibinfo {year} {2015})}\BibitemShut {NoStop}%
\bibitem [{\citenamefont {Li}\ \emph {et~al.}(2015)\citenamefont {Li}, \citenamefont {Ganeshan}, \citenamefont {Pixley},\ and\ \citenamefont {Das~Sarma}}]{li2015manybody}%
  \BibitemOpen
  \bibfield  {author} {\bibinfo {author} {\bibfnamefont {Xiaopeng}\ \bibnamefont {Li}}, \bibinfo {author} {\bibfnamefont {Sriram}\ \bibnamefont {Ganeshan}}, \bibinfo {author} {\bibfnamefont {J.~H.}\ \bibnamefont {Pixley}}, \ and\ \bibinfo {author} {\bibfnamefont {S.}~\bibnamefont {Das~Sarma}},\ }\bibfield  {title} {\enquote {\bibinfo {title} {Many-body localization and quantum nonergodicity in a model with a single-particle mobility edge},}\ }\href {\doibase 10.1103/PhysRevLett.115.186601} {\bibfield  {journal} {\bibinfo  {journal} {Physical Review Letters}\ }\textbf {\bibinfo {volume} {115}},\ \bibinfo {pages} {186601} (\bibinfo {year} {2015})}\BibitemShut {NoStop}%
\bibitem [{\citenamefont {Modak}\ and\ \citenamefont {Mukerjee}(2015)}]{modak2015manybody}%
  \BibitemOpen
  \bibfield  {author} {\bibinfo {author} {\bibfnamefont {Ranjan}\ \bibnamefont {Modak}}\ and\ \bibinfo {author} {\bibfnamefont {Subroto}\ \bibnamefont {Mukerjee}},\ }\bibfield  {title} {\enquote {\bibinfo {title} {Many-body localization in the presence of a single-particle mobility edge},}\ }\href {\doibase 10.1103/PhysRevLett.115.230401} {\bibfield  {journal} {\bibinfo  {journal} {Physical Review Letters}\ }\textbf {\bibinfo {volume} {115}},\ \bibinfo {pages} {230401} (\bibinfo {year} {2015})}\BibitemShut {NoStop}%
\bibitem [{\citenamefont {Li}\ \emph {et~al.}(2016)\citenamefont {Li}, \citenamefont {Pixley}, \citenamefont {Deng}, \citenamefont {Ganeshan},\ and\ \citenamefont {Das~Sarma}}]{li2016quantum}%
  \BibitemOpen
  \bibfield  {author} {\bibinfo {author} {\bibfnamefont {Xiaopeng}\ \bibnamefont {Li}}, \bibinfo {author} {\bibfnamefont {J.~H.}\ \bibnamefont {Pixley}}, \bibinfo {author} {\bibfnamefont {Dong-Ling}\ \bibnamefont {Deng}}, \bibinfo {author} {\bibfnamefont {Sriram}\ \bibnamefont {Ganeshan}}, \ and\ \bibinfo {author} {\bibfnamefont {S.}~\bibnamefont {Das~Sarma}},\ }\bibfield  {title} {\enquote {\bibinfo {title} {Quantum nonergodicity and fermion localization in a system with a single-particle mobility edge},}\ }\href {\doibase 10.1103/PhysRevB.93.184204} {\bibfield  {journal} {\bibinfo  {journal} {Physical Review B}\ }\textbf {\bibinfo {volume} {93}},\ \bibinfo {pages} {184204} (\bibinfo {year} {2016})}\BibitemShut {NoStop}%
\bibitem [{\citenamefont {Deng}\ \emph {et~al.}(2017)\citenamefont {Deng}, \citenamefont {Ganeshan}, \citenamefont {Li}, \citenamefont {Modak}, \citenamefont {Mukerjee},\ and\ \citenamefont {Pixley}}]{deng2017manybody}%
  \BibitemOpen
  \bibfield  {author} {\bibinfo {author} {\bibfnamefont {Dong-Ling}\ \bibnamefont {Deng}}, \bibinfo {author} {\bibfnamefont {Sriram}\ \bibnamefont {Ganeshan}}, \bibinfo {author} {\bibfnamefont {Xiaopeng}\ \bibnamefont {Li}}, \bibinfo {author} {\bibfnamefont {Ranjan}\ \bibnamefont {Modak}}, \bibinfo {author} {\bibfnamefont {Subroto}\ \bibnamefont {Mukerjee}}, \ and\ \bibinfo {author} {\bibfnamefont {J.~H.}\ \bibnamefont {Pixley}},\ }\bibfield  {title} {\enquote {\bibinfo {title} {Many-body localization in incommensurate models with a mobility edge},}\ }\href {\doibase 10.1002/andp.201600399} {\bibfield  {journal} {\bibinfo  {journal} {Annalen der Physik}\ }\textbf {\bibinfo {volume} {529}},\ \bibinfo {pages} {1600399} (\bibinfo {year} {2017})}\BibitemShut {NoStop}%
\bibitem [{\citenamefont {Modak}\ \emph {et~al.}(2018)\citenamefont {Modak}, \citenamefont {Ghosh},\ and\ \citenamefont {Mukerjee}}]{modak2018criterion}%
  \BibitemOpen
  \bibfield  {author} {\bibinfo {author} {\bibfnamefont {Ranjan}\ \bibnamefont {Modak}}, \bibinfo {author} {\bibfnamefont {Soumi}\ \bibnamefont {Ghosh}}, \ and\ \bibinfo {author} {\bibfnamefont {Subroto}\ \bibnamefont {Mukerjee}},\ }\bibfield  {title} {\enquote {\bibinfo {title} {Criterion for the occurrence of many-body localization in the presence of a single-particle mobility edge},}\ }\href {\doibase 10.1103/PhysRevB.97.104204} {\bibfield  {journal} {\bibinfo  {journal} {Physical Review B}\ }\textbf {\bibinfo {volume} {97}},\ \bibinfo {pages} {104204} (\bibinfo {year} {2018})}\BibitemShut {NoStop}%
\bibitem [{\citenamefont {Huang}\ \emph {et~al.}(2023)\citenamefont {Huang}, \citenamefont {Vu}, \citenamefont {Li},\ and\ \citenamefont {Das~Sarma}}]{huang2023incommensurate}%
  \BibitemOpen
  \bibfield  {author} {\bibinfo {author} {\bibfnamefont {Ke}~\bibnamefont {Huang}}, \bibinfo {author} {\bibfnamefont {DinhDuy}\ \bibnamefont {Vu}}, \bibinfo {author} {\bibfnamefont {Xiao}\ \bibnamefont {Li}}, \ and\ \bibinfo {author} {\bibfnamefont {S.}~\bibnamefont {Das~Sarma}},\ }\bibfield  {title} {\enquote {\bibinfo {title} {Incommensurate many-body localization in the presence of long-range hopping and single-particle mobility edge},}\ }\href {\doibase 10.1103/PhysRevB.107.035129} {\bibfield  {journal} {\bibinfo  {journal} {Physical Review B}\ }\textbf {\bibinfo {volume} {107}},\ \bibinfo {pages} {035129} (\bibinfo {year} {2023})}\BibitemShut {NoStop}%
\bibitem [{\citenamefont {Amico}\ \emph {et~al.}(2008)\citenamefont {Amico}, \citenamefont {Fazio}, \citenamefont {Osterloh},\ and\ \citenamefont {Vedral}}]{amico2008entanglement}%
  \BibitemOpen
  \bibfield  {author} {\bibinfo {author} {\bibfnamefont {Luigi}\ \bibnamefont {Amico}}, \bibinfo {author} {\bibfnamefont {Rosario}\ \bibnamefont {Fazio}}, \bibinfo {author} {\bibfnamefont {Andreas}\ \bibnamefont {Osterloh}}, \ and\ \bibinfo {author} {\bibfnamefont {Vlatko}\ \bibnamefont {Vedral}},\ }\bibfield  {title} {\enquote {\bibinfo {title} {Entanglement in many-body systems},}\ }\href {\doibase 10.1103/RevModPhys.80.517} {\bibfield  {journal} {\bibinfo  {journal} {Reviews of Modern Physics}\ }\textbf {\bibinfo {volume} {80}},\ \bibinfo {pages} {517--576} (\bibinfo {year} {2008})}\BibitemShut {NoStop}%
\bibitem [{\citenamefont {Eisert}\ \emph {et~al.}(2010)\citenamefont {Eisert}, \citenamefont {Cramer},\ and\ \citenamefont {Plenio}}]{eisert2010colloquium}%
  \BibitemOpen
  \bibfield  {author} {\bibinfo {author} {\bibfnamefont {J.}~\bibnamefont {Eisert}}, \bibinfo {author} {\bibfnamefont {M.}~\bibnamefont {Cramer}}, \ and\ \bibinfo {author} {\bibfnamefont {M.~B.}\ \bibnamefont {Plenio}},\ }\bibfield  {title} {\enquote {\bibinfo {title} {Colloquium: Area laws for the entanglement entropy},}\ }\href {\doibase 10.1103/RevModPhys.82.277} {\bibfield  {journal} {\bibinfo  {journal} {Reviews of Modern Physics}\ }\textbf {\bibinfo {volume} {82}},\ \bibinfo {pages} {277--306} (\bibinfo {year} {2010})}\BibitemShut {NoStop}%
\bibitem [{\citenamefont {Bauer}\ and\ \citenamefont {Nayak}(2013)}]{bauer2013area}%
  \BibitemOpen
  \bibfield  {author} {\bibinfo {author} {\bibfnamefont {Bela}\ \bibnamefont {Bauer}}\ and\ \bibinfo {author} {\bibfnamefont {Chetan}\ \bibnamefont {Nayak}},\ }\bibfield  {title} {\enquote {\bibinfo {title} {Area laws in a many-body localized state and its implications for topological order},}\ }\href {\doibase 10.1088/1742-5468/2013/09/P09005} {\bibfield  {journal} {\bibinfo  {journal} {Journal of Statistical Mechanics: Theory and Experiment}\ }\textbf {\bibinfo {volume} {2013}},\ \bibinfo {pages} {P09005} (\bibinfo {year} {2013})}\BibitemShut {NoStop}%
\bibitem [{\citenamefont {Kaufman}\ \emph {et~al.}(2016)\citenamefont {Kaufman}, \citenamefont {Tai}, \citenamefont {Lukin}, \citenamefont {Rispoli}, \citenamefont {Schittko}, \citenamefont {Preiss},\ and\ \citenamefont {Greiner}}]{kaufman2016quantum}%
  \BibitemOpen
  \bibfield  {author} {\bibinfo {author} {\bibfnamefont {Adam~M.}\ \bibnamefont {Kaufman}}, \bibinfo {author} {\bibfnamefont {M.~Eric}\ \bibnamefont {Tai}}, \bibinfo {author} {\bibfnamefont {Alexander}\ \bibnamefont {Lukin}}, \bibinfo {author} {\bibfnamefont {Matthew}\ \bibnamefont {Rispoli}}, \bibinfo {author} {\bibfnamefont {Robert}\ \bibnamefont {Schittko}}, \bibinfo {author} {\bibfnamefont {Philipp~M.}\ \bibnamefont {Preiss}}, \ and\ \bibinfo {author} {\bibfnamefont {Markus}\ \bibnamefont {Greiner}},\ }\bibfield  {title} {\enquote {\bibinfo {title} {Quantum thermalization through entanglement in an isolated many-body system},}\ }\href {\doibase 10.1126/science.aaf6725} {\bibfield  {journal} {\bibinfo  {journal} {Science}\ }\textbf {\bibinfo {volume} {353}},\ \bibinfo {pages} {794--800} (\bibinfo {year} {2016})}\BibitemShut {NoStop}%
\bibitem [{\citenamefont {Bardarson}\ \emph {et~al.}(2012)\citenamefont {Bardarson}, \citenamefont {Pollmann},\ and\ \citenamefont {Moore}}]{bardarson2012unbounded}%
  \BibitemOpen
  \bibfield  {author} {\bibinfo {author} {\bibfnamefont {Jens~H.}\ \bibnamefont {Bardarson}}, \bibinfo {author} {\bibfnamefont {Frank}\ \bibnamefont {Pollmann}}, \ and\ \bibinfo {author} {\bibfnamefont {Joel~E.}\ \bibnamefont {Moore}},\ }\bibfield  {title} {\enquote {\bibinfo {title} {Unbounded growth of entanglement in models of many-body localization},}\ }\href {\doibase 10.1103/PhysRevLett.109.017202} {\bibfield  {journal} {\bibinfo  {journal} {Physical Review Letters}\ }\textbf {\bibinfo {volume} {109}},\ \bibinfo {pages} {017202} (\bibinfo {year} {2012})}\BibitemShut {NoStop}%
\bibitem [{\citenamefont {Nanduri}\ \emph {et~al.}(2014)\citenamefont {Nanduri}, \citenamefont {Kim},\ and\ \citenamefont {Huse}}]{nanduri2014entanglement}%
  \BibitemOpen
  \bibfield  {author} {\bibinfo {author} {\bibfnamefont {Arun}\ \bibnamefont {Nanduri}}, \bibinfo {author} {\bibfnamefont {Hyungwon}\ \bibnamefont {Kim}}, \ and\ \bibinfo {author} {\bibfnamefont {David~A.}\ \bibnamefont {Huse}},\ }\bibfield  {title} {\enquote {\bibinfo {title} {Entanglement spreading in a many-body localized system},}\ }\href {\doibase 10.1103/PhysRevB.90.064201} {\bibfield  {journal} {\bibinfo  {journal} {Physical Review B}\ }\textbf {\bibinfo {volume} {90}},\ \bibinfo {pages} {064201} (\bibinfo {year} {2014})}\BibitemShut {NoStop}%
\bibitem [{\citenamefont {{Lewis-Swan}}\ \emph {et~al.}(2019)\citenamefont {{Lewis-Swan}}, \citenamefont {{Safavi-Naini}}, \citenamefont {Kaufman},\ and\ \citenamefont {Rey}}]{lewis-swan2019dynamics}%
  \BibitemOpen
  \bibfield  {author} {\bibinfo {author} {\bibfnamefont {R.~J.}\ \bibnamefont {{Lewis-Swan}}}, \bibinfo {author} {\bibfnamefont {A.}~\bibnamefont {{Safavi-Naini}}}, \bibinfo {author} {\bibfnamefont {A.~M.}\ \bibnamefont {Kaufman}}, \ and\ \bibinfo {author} {\bibfnamefont {A.~M.}\ \bibnamefont {Rey}},\ }\bibfield  {title} {\enquote {\bibinfo {title} {Dynamics of quantum information},}\ }\href {\doibase 10.1038/s42254-019-0090-y} {\bibfield  {journal} {\bibinfo  {journal} {Nature Reviews Physics}\ }\textbf {\bibinfo {volume} {1}},\ \bibinfo {pages} {627--634} (\bibinfo {year} {2019})}\BibitemShut {NoStop}%
\bibitem [{\citenamefont {Wolf}\ \emph {et~al.}(2008)\citenamefont {Wolf}, \citenamefont {Verstraete}, \citenamefont {Hastings},\ and\ \citenamefont {Cirac}}]{wolf2008area}%
  \BibitemOpen
  \bibfield  {author} {\bibinfo {author} {\bibfnamefont {Michael~M.}\ \bibnamefont {Wolf}}, \bibinfo {author} {\bibfnamefont {Frank}\ \bibnamefont {Verstraete}}, \bibinfo {author} {\bibfnamefont {Matthew~B.}\ \bibnamefont {Hastings}}, \ and\ \bibinfo {author} {\bibfnamefont {J.~Ignacio}\ \bibnamefont {Cirac}},\ }\bibfield  {title} {\enquote {\bibinfo {title} {Area laws in quantum systems: Mutual information and correlations},}\ }\href {\doibase 10.1103/PhysRevLett.100.070502} {\bibfield  {journal} {\bibinfo  {journal} {Physical Review Letters}\ }\textbf {\bibinfo {volume} {100}},\ \bibinfo {pages} {070502} (\bibinfo {year} {2008})}\BibitemShut {NoStop}%
\bibitem [{\citenamefont {Hosur}\ \emph {et~al.}(2016)\citenamefont {Hosur}, \citenamefont {Qi}, \citenamefont {Roberts},\ and\ \citenamefont {Yoshida}}]{hosur2016chaos}%
  \BibitemOpen
  \bibfield  {author} {\bibinfo {author} {\bibfnamefont {Pavan}\ \bibnamefont {Hosur}}, \bibinfo {author} {\bibfnamefont {Xiao-Liang}\ \bibnamefont {Qi}}, \bibinfo {author} {\bibfnamefont {Daniel~A.}\ \bibnamefont {Roberts}}, \ and\ \bibinfo {author} {\bibfnamefont {Beni}\ \bibnamefont {Yoshida}},\ }\bibfield  {title} {\enquote {\bibinfo {title} {Chaos in quantum channels},}\ }\href {\doibase 10.1007/JHEP02(2016)004} {\bibfield  {journal} {\bibinfo  {journal} {Journal of High Energy Physics}\ }\textbf {\bibinfo {volume} {2016}},\ \bibinfo {pages} {4} (\bibinfo {year} {2016})}\BibitemShut {NoStop}%
\bibitem [{\citenamefont {Maldacena}\ \emph {et~al.}(2016)\citenamefont {Maldacena}, \citenamefont {Shenker},\ and\ \citenamefont {Stanford}}]{maldacena2016bound}%
  \BibitemOpen
  \bibfield  {author} {\bibinfo {author} {\bibfnamefont {Juan}\ \bibnamefont {Maldacena}}, \bibinfo {author} {\bibfnamefont {Stephen~H.}\ \bibnamefont {Shenker}}, \ and\ \bibinfo {author} {\bibfnamefont {Douglas}\ \bibnamefont {Stanford}},\ }\bibfield  {title} {\enquote {\bibinfo {title} {A bound on chaos},}\ }\href {\doibase 10.1007/JHEP08(2016)106} {\bibfield  {journal} {\bibinfo  {journal} {Journal of High Energy Physics}\ }\textbf {\bibinfo {volume} {2016}},\ \bibinfo {pages} {106} (\bibinfo {year} {2016})}\BibitemShut {NoStop}%
\bibitem [{\citenamefont {Fan}\ \emph {et~al.}(2017)\citenamefont {Fan}, \citenamefont {Zhang}, \citenamefont {Shen},\ and\ \citenamefont {Zhai}}]{fan2017outoftimeorder}%
  \BibitemOpen
  \bibfield  {author} {\bibinfo {author} {\bibfnamefont {Ruihua}\ \bibnamefont {Fan}}, \bibinfo {author} {\bibfnamefont {Pengfei}\ \bibnamefont {Zhang}}, \bibinfo {author} {\bibfnamefont {Huitao}\ \bibnamefont {Shen}}, \ and\ \bibinfo {author} {\bibfnamefont {Hui}\ \bibnamefont {Zhai}},\ }\bibfield  {title} {\enquote {\bibinfo {title} {Out-of-time-order correlation for many-body localization},}\ }\href {\doibase 10.1016/j.scib.2017.04.011} {\bibfield  {journal} {\bibinfo  {journal} {Science Bulletin}\ }\textbf {\bibinfo {volume} {62}},\ \bibinfo {pages} {707--711} (\bibinfo {year} {2017})}\BibitemShut {NoStop}%
\bibitem [{\citenamefont {Nahum}\ \emph {et~al.}(2018)\citenamefont {Nahum}, \citenamefont {Vijay},\ and\ \citenamefont {Haah}}]{nahum2018operator}%
  \BibitemOpen
  \bibfield  {author} {\bibinfo {author} {\bibfnamefont {Adam}\ \bibnamefont {Nahum}}, \bibinfo {author} {\bibfnamefont {Sagar}\ \bibnamefont {Vijay}}, \ and\ \bibinfo {author} {\bibfnamefont {Jeongwan}\ \bibnamefont {Haah}},\ }\bibfield  {title} {\enquote {\bibinfo {title} {Operator spreading in random unitary circuits},}\ }\href {\doibase 10.1103/PhysRevX.8.021014} {\bibfield  {journal} {\bibinfo  {journal} {Physical Review X}\ }\textbf {\bibinfo {volume} {8}},\ \bibinfo {pages} {021014} (\bibinfo {year} {2018})}\BibitemShut {NoStop}%
\bibitem [{\citenamefont {Chen}\ \emph {et~al.}(2025)\citenamefont {Chen}, \citenamefont {Liu},\ and\ \citenamefont {Zhang}}]{chen2025subsystem}%
  \BibitemOpen
  \bibfield  {author} {\bibinfo {author} {\bibfnamefont {Yu-Qin}\ \bibnamefont {Chen}}, \bibinfo {author} {\bibfnamefont {Shuo}\ \bibnamefont {Liu}}, \ and\ \bibinfo {author} {\bibfnamefont {Shi-Xin}\ \bibnamefont {Zhang}},\ }\bibfield  {title} {\enquote {\bibinfo {title} {Subsystem information capacity in random circuits and hamiltonian dynamics},}\ }\href {\doibase 10.22331/q-2025-06-24-1783} {\bibfield  {journal} {\bibinfo  {journal} {Quantum}\ }\textbf {\bibinfo {volume} {9}},\ \bibinfo {pages} {1783} (\bibinfo {year} {2025})}\BibitemShut {NoStop}%
\bibitem [{\citenamefont {Holevo}(2012)}]{holevo2012quantum}%
  \BibitemOpen
  \bibfield  {author} {\bibinfo {author} {\bibfnamefont {Alexander~S.}\ \bibnamefont {Holevo}},\ }\href {\doibase 10.1515/9783110273403} {\emph {\bibinfo {title} {Quantum Systems, Channels, Information: A Mathematical Introduction}}}\ (\bibinfo  {publisher} {De Gruyter},\ \bibinfo {year} {2012})\BibitemShut {NoStop}%
\bibitem [{\citenamefont {Schumacher}\ and\ \citenamefont {Nielsen}(1996)}]{schumacher1996quantum}%
  \BibitemOpen
  \bibfield  {author} {\bibinfo {author} {\bibfnamefont {Benjamin}\ \bibnamefont {Schumacher}}\ and\ \bibinfo {author} {\bibfnamefont {M.~A.}\ \bibnamefont {Nielsen}},\ }\bibfield  {title} {\enquote {\bibinfo {title} {Quantum data processing and error correction},}\ }\href {\doibase 10.1103/PhysRevA.54.2629} {\bibfield  {journal} {\bibinfo  {journal} {Physical Review A}\ }\textbf {\bibinfo {volume} {54}},\ \bibinfo {pages} {2629--2635} (\bibinfo {year} {1996})}\BibitemShut {NoStop}%
\bibitem [{\citenamefont {Lloyd}(1997)}]{lloyd1997capacity}%
  \BibitemOpen
  \bibfield  {author} {\bibinfo {author} {\bibfnamefont {Seth}\ \bibnamefont {Lloyd}},\ }\bibfield  {title} {\enquote {\bibinfo {title} {Capacity of the noisy quantum channel},}\ }\href {\doibase 10.1103/PhysRevA.55.1613} {\bibfield  {journal} {\bibinfo  {journal} {Physical Review A}\ }\textbf {\bibinfo {volume} {55}},\ \bibinfo {pages} {1613--1622} (\bibinfo {year} {1997})}\BibitemShut {NoStop}%
\bibitem [{\citenamefont {Devetak}\ and\ \citenamefont {Winter}(2005)}]{devetak2005distillation}%
  \BibitemOpen
  \bibfield  {author} {\bibinfo {author} {\bibfnamefont {Igor}\ \bibnamefont {Devetak}}\ and\ \bibinfo {author} {\bibfnamefont {Andreas}\ \bibnamefont {Winter}},\ }\bibfield  {title} {\enquote {\bibinfo {title} {Distillation of secret key and entanglement from quantum states},}\ }\href {\doibase 10.1098/rspa.2004.1372} {\bibfield  {journal} {\bibinfo  {journal} {Proceedings of the Royal Society A: Mathematical, Physical and Engineering Sciences}\ }\textbf {\bibinfo {volume} {461}},\ \bibinfo {pages} {207--235} (\bibinfo {year} {2005})}\BibitemShut {NoStop}%
\bibitem [{\citenamefont {Zhang}\ \emph {et~al.}(2023)\citenamefont {Zhang}, \citenamefont {Allcock}, \citenamefont {Wan}, \citenamefont {Liu}, \citenamefont {Sun}, \citenamefont {Yu}, \citenamefont {Yang}, \citenamefont {Qiu}, \citenamefont {Ye}, \citenamefont {Chen}, \citenamefont {Lee}, \citenamefont {Zheng}, \citenamefont {Jian}, \citenamefont {Yao}, \citenamefont {Hsieh},\ and\ \citenamefont {Zhang}}]{Zhang2023tensorcircuit}%
  \BibitemOpen
  \bibfield  {author} {\bibinfo {author} {\bibfnamefont {Shi-Xin}\ \bibnamefont {Zhang}}, \bibinfo {author} {\bibfnamefont {Jonathan}\ \bibnamefont {Allcock}}, \bibinfo {author} {\bibfnamefont {Zhou-Quan}\ \bibnamefont {Wan}}, \bibinfo {author} {\bibfnamefont {Shuo}\ \bibnamefont {Liu}}, \bibinfo {author} {\bibfnamefont {Jiace}\ \bibnamefont {Sun}}, \bibinfo {author} {\bibfnamefont {Hao}\ \bibnamefont {Yu}}, \bibinfo {author} {\bibfnamefont {Xing-Han}\ \bibnamefont {Yang}}, \bibinfo {author} {\bibfnamefont {Jiezhong}\ \bibnamefont {Qiu}}, \bibinfo {author} {\bibfnamefont {Zhaofeng}\ \bibnamefont {Ye}}, \bibinfo {author} {\bibfnamefont {Yu-Qin}\ \bibnamefont {Chen}}, \bibinfo {author} {\bibfnamefont {Chee-Kong}\ \bibnamefont {Lee}}, \bibinfo {author} {\bibfnamefont {Yi-Cong}\ \bibnamefont {Zheng}}, \bibinfo {author} {\bibfnamefont {Shao-Kai}\ \bibnamefont {Jian}}, \bibinfo {author} {\bibfnamefont {Hong}\ \bibnamefont {Yao}}, \bibinfo {author} {\bibfnamefont {Chang-Yu}\ \bibnamefont {Hsieh}}, \ and\ \bibinfo {author} {\bibfnamefont {Shengyu}\ \bibnamefont {Zhang}},\ }\bibfield  {title} {\enquote {\bibinfo {title} {Tensorcircuit: a quantum software framework for the nisq era},}\ }\href {\doibase 10.22331/q-2023-02-02-912} {\bibfield  {journal} {\bibinfo  {journal} {Quantum}\ }\textbf {\bibinfo {volume} {7}},\ \bibinfo {pages} {912} (\bibinfo {year} {2023})}\BibitemShut {NoStop}%
\bibitem [{sm(2025)}]{sm}%
  \BibitemOpen
  \href@noop {} {\enquote {\bibinfo {title} {See supplemental material at [url] for details on the extraction of entanglement metrics, results from domain-wall and random initial states, and robustness checks for sic under periodic boundary conditions},}\ } (\bibinfo {year} {2025})\BibitemShut {NoStop}%
\bibitem [{\citenamefont {Longhi}(2021)}]{longhi2021phase}%
  \BibitemOpen
  \bibfield  {author} {\bibinfo {author} {\bibfnamefont {Stefano}\ \bibnamefont {Longhi}},\ }\bibfield  {title} {\enquote {\bibinfo {title} {Phase transitions in a non-hermitian aubry-andr\textbackslash 'e-harper model},}\ }\href {\doibase 10.1103/PhysRevB.103.054203} {\bibfield  {journal} {\bibinfo  {journal} {Physical Review B}\ }\textbf {\bibinfo {volume} {103}},\ \bibinfo {pages} {054203} (\bibinfo {year} {2021})}\BibitemShut {NoStop}%
\bibitem [{\citenamefont {Shimasaki}\ \emph {et~al.}(2024)\citenamefont {Shimasaki}, \citenamefont {Bai}, \citenamefont {Kondakci}, \citenamefont {Dotti}, \citenamefont {Pagett}, \citenamefont {Dardia}, \citenamefont {Prichard}, \citenamefont {Eckardt},\ and\ \citenamefont {Weld}}]{shimasaki2024reversible}%
  \BibitemOpen
  \bibfield  {author} {\bibinfo {author} {\bibfnamefont {Toshihiko}\ \bibnamefont {Shimasaki}}, \bibinfo {author} {\bibfnamefont {Yifei}\ \bibnamefont {Bai}}, \bibinfo {author} {\bibfnamefont {H.~Esat}\ \bibnamefont {Kondakci}}, \bibinfo {author} {\bibfnamefont {Peter}\ \bibnamefont {Dotti}}, \bibinfo {author} {\bibfnamefont {Jared~E.}\ \bibnamefont {Pagett}}, \bibinfo {author} {\bibfnamefont {Anna~R.}\ \bibnamefont {Dardia}}, \bibinfo {author} {\bibfnamefont {Max}\ \bibnamefont {Prichard}}, \bibinfo {author} {\bibfnamefont {Andr{\'e}}\ \bibnamefont {Eckardt}}, \ and\ \bibinfo {author} {\bibfnamefont {David~M.}\ \bibnamefont {Weld}},\ }\bibfield  {title} {\enquote {\bibinfo {title} {Reversible phasonic control of a quantum phase transition in a quasicrystal},}\ }\href {\doibase 10.1103/PhysRevLett.133.083405} {\bibfield  {journal} {\bibinfo  {journal} {Physical Review Letters}\ }\textbf {\bibinfo {volume} {133}},\ \bibinfo {pages} {083405} (\bibinfo {year} {2024})}\BibitemShut {NoStop}%
\bibitem [{Note1()}]{Note1}%
  \BibitemOpen
  \bibinfo {note} {\protect \url {https://gitee.com/yuqiqing/sic_in_-gaa_model}}\BibitemShut {NoStop}%
\end{thebibliography}%


\begin{thebibliography}{0}%
\makeatletter
\providecommand \@ifxundefined [1]{%
 \@ifx{#1\undefined}
}%
\providecommand \@ifnum [1]{%
 \ifnum #1\expandafter \@firstoftwo
 \else \expandafter \@secondoftwo
 \fi
}%
\providecommand \@ifx [1]{%
 \ifx #1\expandafter \@firstoftwo
 \else \expandafter \@secondoftwo
 \fi
}%
\providecommand \natexlab [1]{#1}%
\providecommand \enquote  [1]{``#1''}%
\providecommand \bibnamefont  [1]{#1}%
\providecommand \bibfnamefont [1]{#1}%
\providecommand \citenamefont [1]{#1}%
\providecommand \href@noop [0]{\@secondoftwo}%
\providecommand \href [0]{\begingroup \@sanitize@url \@href}%
\providecommand \@href[1]{\@@startlink{#1}\@@href}%
\providecommand \@@href[1]{\endgroup#1\@@endlink}%
\providecommand \@sanitize@url [0]{\catcode `\\12\catcode `\$12\catcode `\&12\catcode `\#12\catcode `\^12\catcode `\_12\catcode `\%12\relax}%
\providecommand \@@startlink[1]{}%
\providecommand \@@endlink[0]{}%
\providecommand \url  [0]{\begingroup\@sanitize@url \@url }%
\providecommand \@url [1]{\endgroup\@href {#1}{\urlprefix }}%
\providecommand \urlprefix  [0]{URL }%
\providecommand \Eprint [0]{\href }%
\providecommand \doibase [0]{http://dx.doi.org/}%
\providecommand \selectlanguage [0]{\@gobble}%
\providecommand \bibinfo  [0]{\@secondoftwo}%
\providecommand \bibfield  [0]{\@secondoftwo}%
\providecommand \translation [1]{[#1]}%
\providecommand \BibitemOpen [0]{}%
\providecommand \bibitemStop [0]{}%
\providecommand \bibitemNoStop [0]{.\EOS\space}%
\providecommand \EOS [0]{\spacefactor3000\relax}%
\providecommand \BibitemShut  [1]{\csname bibitem#1\endcsname}%
\let\auto@bib@innerbib\@empty
\end{thebibliography}%
\let\addcontentsline\oldaddcontentsline

\end{document}


\begin{center}
\textbf{\large Supplemental Material for ``Entanglement growth and information capacity in a quasiperiodic system with a single-particle mobility edge''}
\end{center}

\author{Yuqi Qing}
\affiliation{Institute of Physics, Chinese Academy of Sciences, Beijing 100190, China}
\affiliation{Tsung-Dao Lee Institute, Shanghai Jiao Tong University, Shanghai 201210, China}

\author{Yu-Qin Chen}
\email{yqchen@gscaep.ac.cn}
\affiliation{Graduate School of China Academy of Engineering Physics, Beijing 100193, China}

\author{Shi-Xin Zhang}
\email{shixinzhang@iphy.ac.cn}
\affiliation{Institute of Physics, Chinese Academy of Sciences, Beijing 100190, China}

\date{\today}

\maketitle

\renewcommand{\thefigure}{S\arabic{figure}}
\setcounter{figure}{0}
\renewcommand{\theequation}{S\arabic{equation}}
\setcounter{equation}{0}
\renewcommand{\thesection}{\Roman{section}}
\setcounter{section}{0}
\setcounter{secnumdepth}{4}


\section{Time evolution for entanglement entropy}

\begin{figure}[h]
    \centering
    \includegraphics[width=1\textwidth]{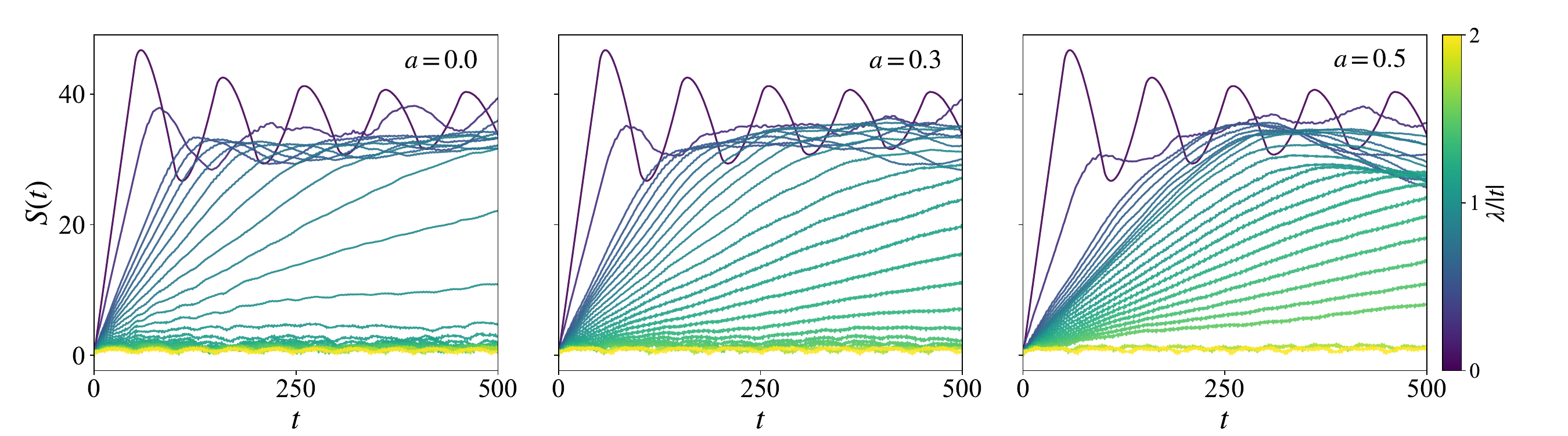}
    \caption{Time evolution of half-chain EE from the N\'eel state. The plots show EE as a function of time \(t\) for different values of the potential strength \(\lambda/|t|\) (from \(\lambda/|t|=0\), top curve, to \(\lambda/|t|=2.0\), bottom curve) and deformation parameter \(a\). The system size is \(L=200\). The dynamics clearly show an initial growth phase followed by the saturation stage.}
    \label{figS1}
\end{figure}

In this section, we provide details on the extraction of the early-time entanglement growth velocity \(v_S\) and the late-time saturation value \(S_{\mathrm{sat}}\) from the time evolution of the half-chain entanglement entropy (EE). Fig.~\ref{figS1} shows the time evolution of EE for a quench from the N\'eel state. The dynamics can be separated into two distinct regimes:

\begin{enumerate}
    \item \textbf{Early-time linear growth regime:} For short times after the quench, the EE grows approximately linearly. We extract the velocity \(v_S\) by performing a linear fit to the EE curve in this initial period. In practice, the time window \(t \in [0, 20]\) serves as a reasonable choice to identify the early-time regime in the system under investigation. 
    \item \textbf{Late-time saturation regime:} At long times, the EE stops growing and oscillates around a steady-state value. We calculate the saturation value \(S_{\mathrm{sat}}\) by averaging the EE over a long-time window in this saturation regime, after the initial growth has ceased. In our calculation, we first quench the system for a sufficiently long time \(t = 10000\), and then take 1000 samples with an average time interval of \(\langle dt \rangle = 10\). To ensure adequate sampling along the EE profile, the interval \(dt\) is randomly varied around the average.
\end{enumerate}

The values of \(v_S\) and \(S_{\mathrm{sat}}\) presented in the main text and this supplemental material are obtained using this procedure.


\section{Entanglement growth from other initial states}

To ensure the generality of our conclusions, we supplement the results from the N\'eel state with data from two other types of initial product states at half-filling, the domain-wall state (\(|11\dots100\dots0\rangle\)) and random product states. For the random states, we average the results over different random configurations of \(L/2\) fermions.

\begin{figure}[h]
    \centering
    \includegraphics[width=1\textwidth]{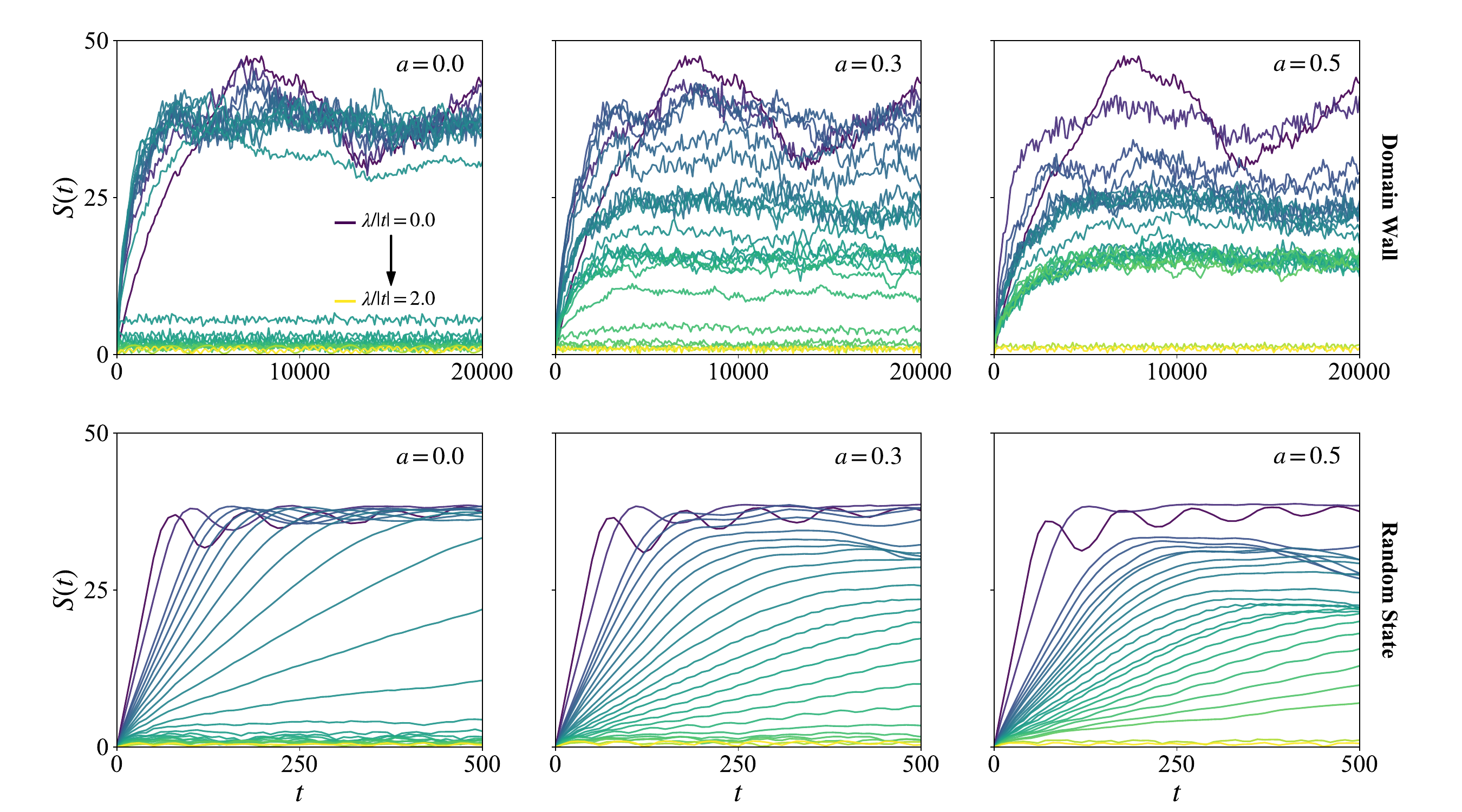}
    \caption{Time evolution of EE for other initial states. (Top row) Quench from a domain-wall state. (Bottom row) Quench from random product states (averaged). The parameters are \(L=200\). The qualitative features of EE growth and saturation are consistent across different initial states.}
    \label{figS2}
\end{figure}

\begin{figure}[h]
    \centering
    \includegraphics[width=0.8\textwidth]{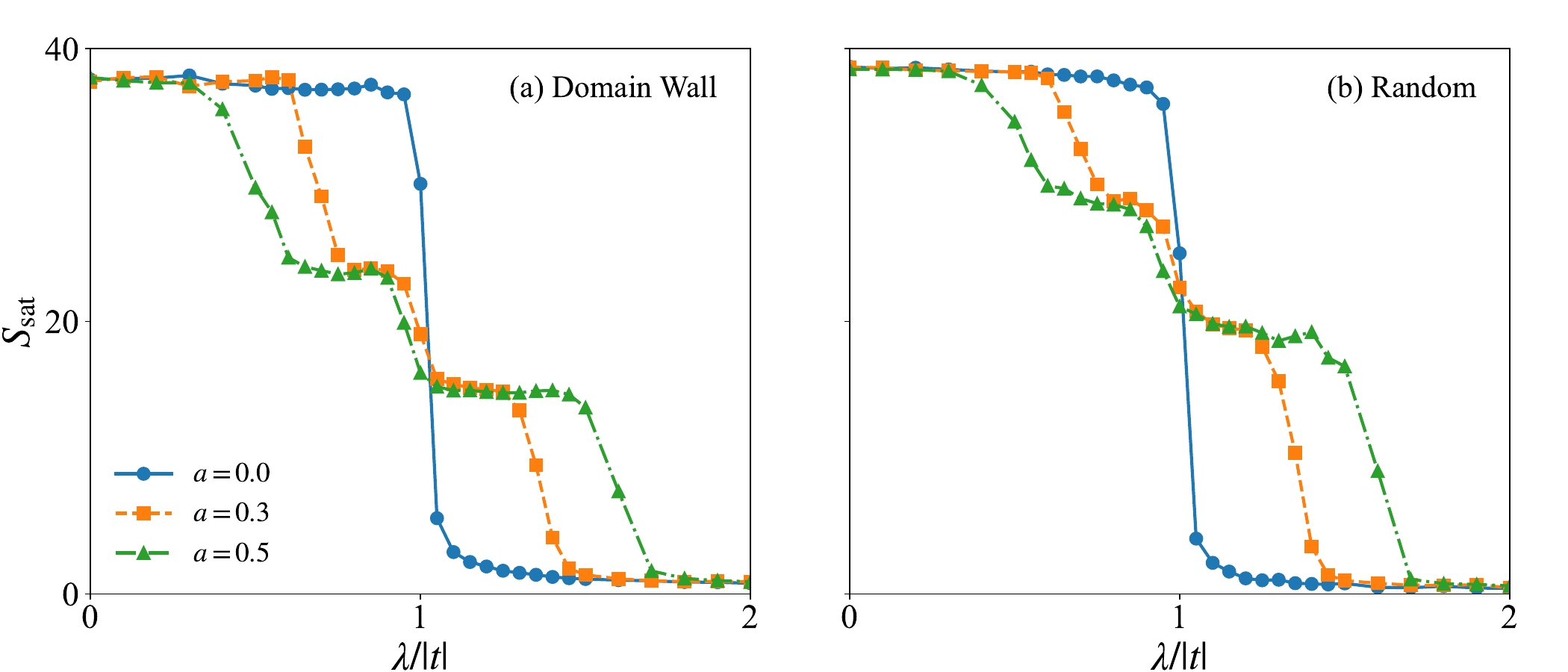}
    \caption{Saturation EE for other initial states. (a) Domain-wall initial state. (b) Random initial states (averaged). The plots show \(S_{\mathrm{sat}}\) versus \(\lambda/|t|\). The smooth crossover in the GAA model (\(a>0\)) with plateaus corresponding to the spectrum gap is a robust feature.}
    \label{figS3}
\end{figure}

\begin{figure}[h]
    \centering
    \includegraphics[width=0.9\textwidth]{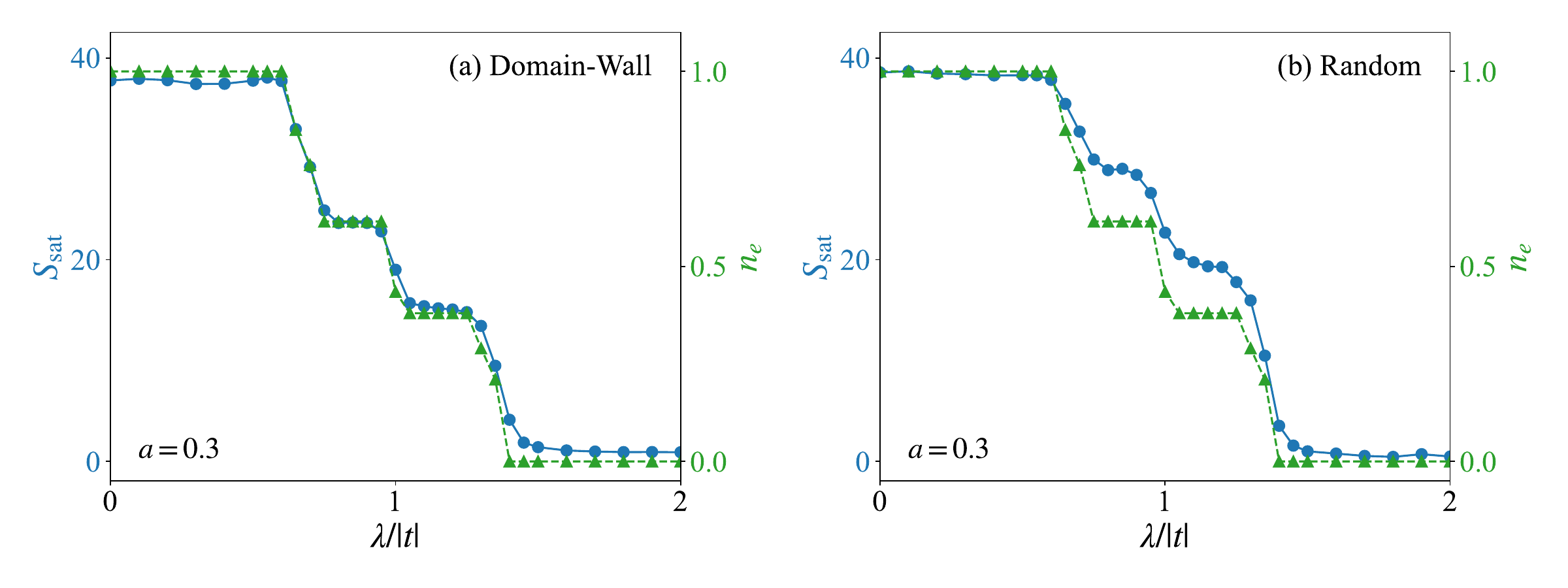}
    \caption{Connection between saturation EE and the number of extended states for other initial states. (a) Domain-wall initial state. (b) Random initial states (averaged). Both plots are for \(a=0.3\) and \(L=200\). \(S_{\mathrm{sat}}\) (blue circles, left axis) strongly correlates with \(N_e\) (green triangles, right axis), confirming the conclusion from the main text.}
    \label{figS4}
\end{figure}

Fig.\ref{figS2} shows the time evolution of EE for these states. While the qualitative behavior is similar to the N\'eel state, we note that for the domain-wall state, the oscillation period in the saturation regime is significantly longer. The saturation values (Fig.~\ref{figS3}) also show trends consistent with the main text, namely a sharp transition for the AA model (\(a=0\)) and a smooth crossover for the GAA model (\(a>0\)). Fig.~\ref{figS4} demonstrates that for both the domain-wall and random initial states, the saturation EE \(S_{\mathrm{sat}}\) is strongly correlated with the fraction of extended states \(n_e\), reinforcing the universality of our central conclusion.


\section{Steady-state SIC profile for edge-coupling and other initial states}

\begin{figure}[h]
    \centering
    \includegraphics[width=1\textwidth]{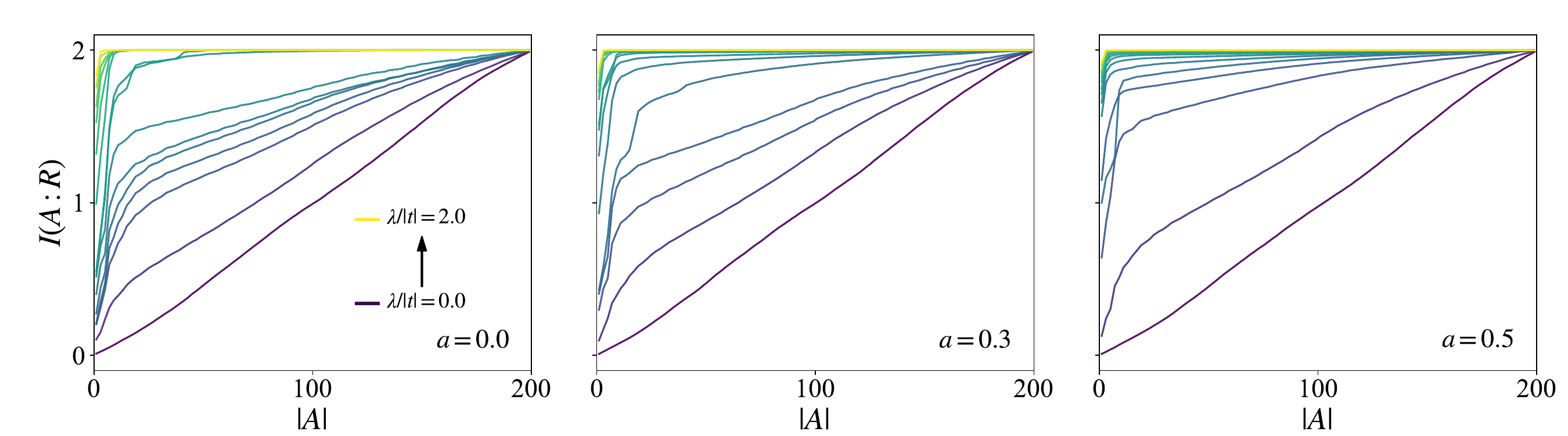}
    \caption{Steady-state SIC profile \(I(A:R)\) for the edge-coupling case. The initial state is N\'eel state. The reference qubit is entangled with the edge site \(E=1\), and the subsystem \(A\) starts from this edge. }
    \label{figS5}
\end{figure}

To study boundary effects, we also investigate the steady-state subsystem information capacity (SIC) profile by coupling the reference site to an edge site (\(E=1\)). As shown in Fig.~\ref{figS5}, the SIC profile for this edge-coupling setup is different from the center-coupling case, especially for the standard AA model. In this case, the SIC profile fails to act as a clear probe for the phase transition, as the profile varies smoothly with increasing $\lambda$ even in the extended phase. This failure might be due to the asymmetric information propagation with boundary reflections and interference.

\begin{figure}[h]
    \centering
    \includegraphics[width=1\textwidth]{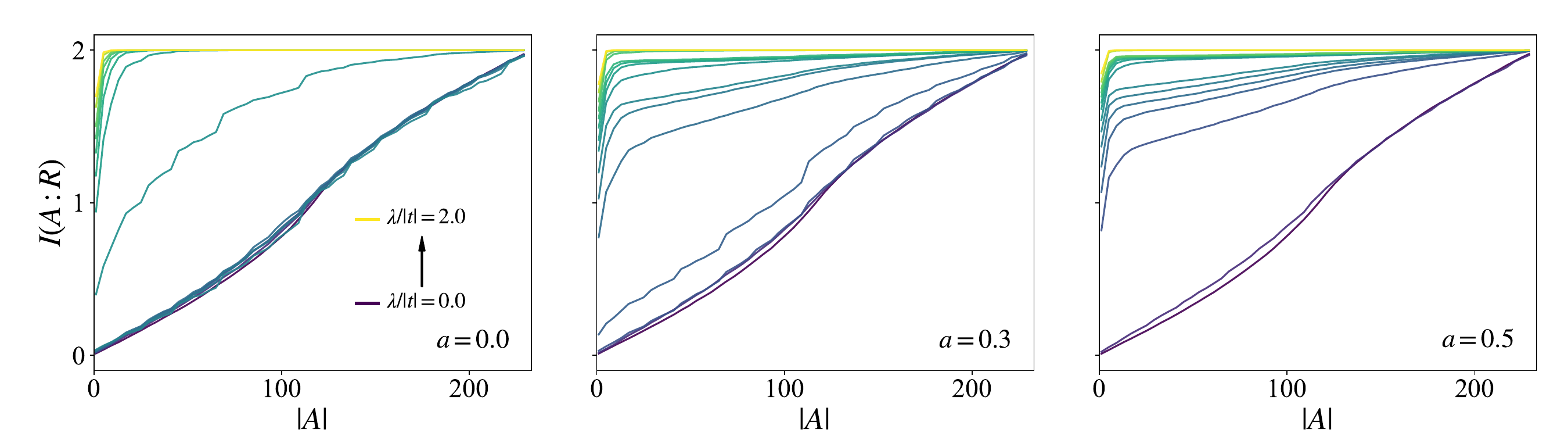}
    \caption{Steady-state SIC profile \(I(A:R)\) for edge coupling under PBC. The system size is \(L=233\) with \(b=144/233\). Under PBC, the edge probe clearly resolves the phase transition and SPME features.}
    \label{figS6}
\end{figure}

To further clarify the role of boundary conditions, we compute the SIC under periodic boundary conditions (PBC) using a rational approximation of the quasiperiodic parameter ($L=233, b=144/233$). As shown in Fig.~\ref{figS6}, eliminating boundary reflections restores the sharp transition for the AA model ($a=0$) and clearly exhibits the hybrid jump-ramp profiles for the GAA model ($a>0$). This confirms that the behavior in Fig.~\ref{figS5} is due to boundary effects in open boundaries.

\begin{figure}[h]
    \centering
    \includegraphics[width=0.95\textwidth]{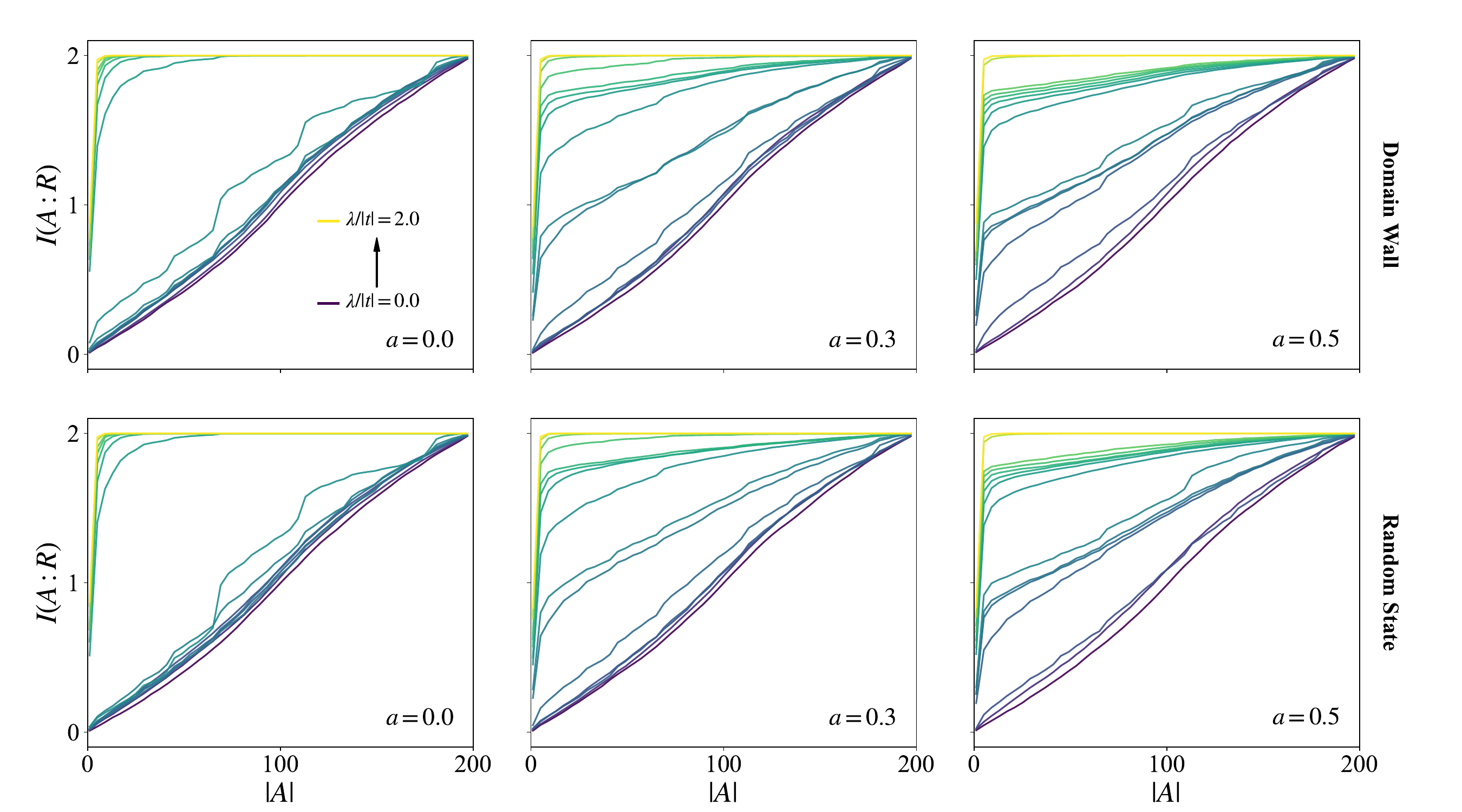}
    \caption{SIC profiles for center-coupling with other initial states. The reference qubit is entangled with the center site \(E=L/2\). (Top row) Domain-wall initial state. (Bottom row) Random initial states (averaged). System size \(L=200\), subsystem size \(|A|\) is measured for the central region. The mixed SIC profile in the GAA model is a generic feature.}
    \label{figS7}
\end{figure}

\begin{figure}[h]
    \centering
    \includegraphics[width=0.95\textwidth]{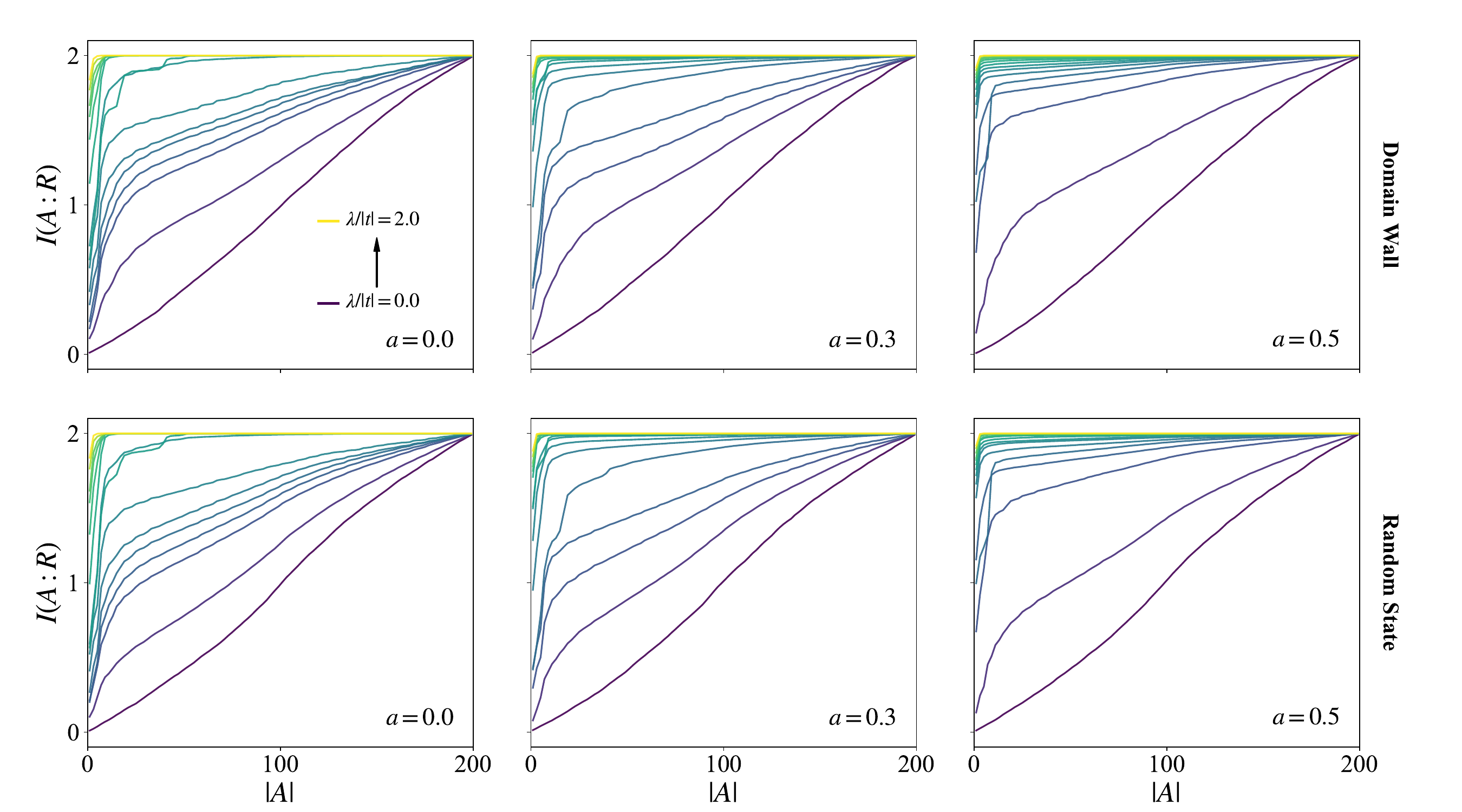}
    \caption{SIC profiles for edge-coupling with other initial states. The reference qubit is entangled with the edge site \(E=1\). (Top row) Domain-wall initial state. (Bottom row) Random initial states (averaged). System size \(L=200\). The results are consistent with those in the main text.}
    \label{figS8}
\end{figure}

Finally, we present the SIC results for different initial states mentioned before. Figures~\ref{figS7} and \ref{figS8} show the SIC profiles for center- and edge-coupling, respectively. The results confirm that the hybrid SIC profile is a generic dynamical signature of the SPME, independent of the specific initial product state.
